\documentclass[
 aip,jcp,
 amsmath,amssymb,
 reprint
]{revtex4-2}

\usepackage{haimi}
\usepackage[T1]{fontenc}
\usepackage{mathptmx}
\usepackage{etoolbox}
\usepackage{physics}
\usepackage{nicefrac}
\usepackage{empheq}
\usepackage{xcolor}
\usepackage[normalem]{ulem}
\newboolean{showannotations}   
\setboolean{showannotations}{false} 
\newcommand*{\added}[1]{%
  \ifthenelse{\boolean{showannotations}}{{\color{blue}#1}}{#1}%
}

\newcommand*{\removed}[1]{%
  \ifthenelse{\boolean{showannotations}}{\st{#1}}{}%
}

\newcommand*{\change}[2]{%
  \ifthenelse{\boolean{showannotations}}{{\color{red}\st{#1}}{\color{blue}#2}}{#2}%
}

\newcommand{\expecth}[3]{\langle#1|#2|#3\rangle}

\begin{document} 

\title{Mixed Quantum-Classical Methods for Polaron Spectral Functions}
\author{Haimi Nguyen}%
\affiliation{Department of Chemistry, Columbia University, 3000 Broadway, New York, New York, 10027,  U.S.A}
\author{Arkajit Mandal}%
\affiliation{Department of Chemistry, Texas A\&M University, College Station, Texas 77843, USA}
\author{Ankit Mahajan}%
\affiliation{Department of Chemistry, Columbia University, 3000 Broadway, New York, New York, 10027,  U.S.A}
\author{David R. Reichman}
\email{drr2103@columbia.edu}
\affiliation{Department of Chemistry, Columbia University, 3000 Broadway, New York, New York, 10027,  U.S.A}

\begin{abstract}
In this work, using two distinct semiclassical approaches, namely the mean field Ehrenfest (MFE) method and the mapping approach to surface hopping (MASH), we investigate the spectral function of a single charge interacting with phonons on a lattice.  This quantity is relevant for the description of angle-resolved photoemission experiments.  Focusing on the one-dimensional Holstein model, we compare the performance of these approaches across a range of hopping parameters, coupling strengths, and lattice sizes at low and high temperatures, exposing the relative strengths and weaknesses of each.  We demonstrate that these approaches can be efficiently applied with reasonable accuracy to {\em ab initio} polaron models.  Our work provides a route to the calculation of spectral properties in realistic electron-phonon-coupled systems in a computationally inexpensive manner with encouraging accuracy.
\end{abstract}

\maketitle
\section{Introduction} 

Electron-phonon interactions (EPI) are of crucial importance in condensed matter physics, playing a critical role in phenomena such as superconductivity, charge transport, and the thermodynamic properties of solids.~\cite{ShirotaCR2007, ashcroft1976solid, HannewaldPRB2004, TroisiPRL2006, CoropceanuCR2007} Prominent models of EPI are the Holstein, Fröhlich, and Su-Schrieffer-Heeger models.~\cite{mahan2013many, XingPRL2021, WangJCP2011} The accurate computation of spectral functions in electron-phonon coupled systems provides essential information such as the density of states and the lifetimes of quasi-particles, and is thus crucial for the interpretation of experimental data obtained from techniques such as angle-resolved photoemission spectroscopy.~\cite{mahan2013many} Among the models listed above, the Holstein model stands out for its simplicity, making it an ideal testing ground for numerical methods while still capturing the essential lattice level features of small polaron formation in systems such as molecular crystals.~\cite{mahan2013many} Calculating the dynamical behavior of the Holstein model is thus important both for uncovering the underlying physics of real materials and for evaluating the effectiveness of different numerical techniques.

Numerically exact spectral functions have been obtained for the Holstein model using a variety of methods. Bon\v{c}a \textit{et al.}~\cite{Bona2019} have employed the finite-temperature Lanczos approach to compute the spectral function for the Holstein polaron on a one-dimensional ring. 
Mitrić \textit{et al.}~\cite{Mitri2022} demonstrated that dynamical mean field theory provides an excellent approximation for spectral functions across all parameter regimes, even in one-dimensional systems where it might be expected to perform poorly. The Hierarchical Equations of Motion (HEOM) technique has also been employed to study spectral properties in a highly accurate manner \cite{Jankovi2022} despite challenges associated with numerical instabilities~\cite{Dunn2019}. 
Other controlled approaches, such as the density matrix renormalization group at zero~\cite{Jeckelmann1998,Zhang1999} and finite~\cite{Jansen2020} temperatures, and the generalized Green's function cluster expansion~\cite{Carbone2021}, have been successfully applied to this problem.
In addition, recent studies~\cite{Mitri2023,Robinson2022} have investigated in detail the validity of the second-order cumulant expansion for calculating spectral functions and quasiparticle properties within the Holstein model, further enriching the methodological toolkit for this and related problems.

Mixed quantum-classical (MQC) methods, where electrons are treated quantum mechanically and phonons are treated classically, offer a promising alternative to these methods.~\cite{Herman1994,Thoss2004} Indeed, a recent study by Mitrić \textit{et al.}~\cite{Mitri2025} shows a remarkable agreement between MQC and exact treatments of the frequency-dependent mobility in the one-dimensional Holstein model near adiabatic limit.
MQC approaches, albeit approximate, offer the flexibility of treating systems beyond simple examples such as the Holstein model, including those where the EPI is non-linear and where the phonons are anharmonic. 
In addition, MQC methods can be used to efficiently simulate the behavior of \textit{ab initio} models of EPI in solids where near-exact approaches become extremely expensive. For example, they have been applied to study charge transport in crystalline rubrene.~\cite{Runeson2024b}
However, there are many variants of MQC techniques, and to date there has been no systematic comparison of MQC in models of EPI to assess their efficacy and accuracy for quantities such as the spectral function.

One of the most widely used MQC methods is the mean field Ehrenfest method (MFE), in which the classical bath dynamically evolves on an average potential surface weighted by electronic wavefunction coefficients. However, it is known to suffer from the issue of overcoherence and incorrect detailed balance, where heating to infinite temperature can occur in the long time limit.~\cite{Parandekar2006,Akimov2014}
Another popular MQC approach is a class of surface hopping methods inspired by Tully's fewest switches surface hopping (FSSH)~\cite{Tully1990,Sholl1998} developed in the 1990s. Unlike MFE, the interaction between the classical bath and the quantum system in FSSH is chosen stochastically among the many potential surfaces, allowing for wave-packet branching that is absent in MFE. This approach is very popular in nonadiabatic dynamics simulations due to its ease of use in \textit{ab initio} settings. Recently, Runeson and Manolopoulos have developed the multi-state mapping approach to surface hopping (MASH)~\cite{Runeson2023,Runeson2024} based on a previous two-state MASH formulation by Mannouch and Richardson~\cite{Mannouch2023,Richardson2025}. Unlike FSSH, MASH selects the potential surface from which to evolve the classical degrees of freedom deterministically. MASH has been shown to obey detailed balance and provide accurate populations in various models.~\cite{Runeson2023,Runeson2024}

In this work, we employ MFE and MASH to calculate the spectral function of the Holstein model across various parameter regimes and compare the results to exact benchmark data. By applying these MQC methods, we can assess the ability of these distinct approaches to reproduce the detailed underlying features encoded in the spectral function. 
Having determined the utility of MFE for the Holstein model, we turn to a more challenging goal where semiclassical methods may be of great utility, namely for the simulation of \textit{ab intio} problems.
Specifically, we utilize MFE to compute the spectral function of the hole polaron in LiF, an archetypal strong-coupling polaron system. 

This paper is organized as follows: In Sec.~\ref{sec:theory}, we describe the models studied in this work, namely the Holstein model and the \textit{ab initio} LiF Hamiltonian. Sec.~\ref{sec:theorymfe} and Sec.~\ref{sec:theorymash} provide a detailed overview of the MFE and MASH methods respectively and explain how spectral functions are computed within these frameworks. Specifically, Sec.~\ref{sec:theoryanalytical_J=0T=0} discusses analytical expressions for the exact, MFE, and MASH spectral functions in the limit of zero hopping amplitude and temperature of the Holstein model. In Sec.~\ref{sec:results}, we present a comparative analysis, first evaluating the performance of MFE against MASH in various regimes, then comparing different variations of MFE and the cumulant expansion method, and finally applying MFE to LiF. We conclude in Sec.~\ref{sec:summary} with a discussion on the utility of MQC methods for calculating spectral functions in the Holstein model and LiF.

\section{Models and Methods}\label{sec:theory} 
In this section, we begin by describing the model systems studied in this work: the one-dimensional periodic Holstein model and an \textit{ab initio} model of a hole in the valence bands of solid LiF. We then proceed to define the main observable we calculate, namely the spectral function, and outline the MFE and MASH algorithms. Finally, we provide analytical expressions for the spectral function of the Holstein model in the analytically solvable limit of zero hopping as a means to gain intuition for our numerical results.
\subsection{Model System} 
\subsubsection{Holstein model}
In this study, we focus on the one-dimensional periodic Holstein model. In real space, the Hamiltonian ($\hbar=1$) is expressed as
\begin{align}
    H
     &=\sum_{n=0}^{N-1} \left[ -J(\hat{c}^\dagger_{n+1}\hat{c}_n + h.c.)+ g\hat{c}^\dagger_{n}\hat{c}_n (\hat{b}^\dagger_n + \hat{b}_n) + \omega_0 \hat{b}^\dagger_n \hat{b}_n \right]. \label{eqn:realspace_Holstein_Hamiltonian}
\end{align}
Periodic boundary conditions are used such that site $N$ is equivalent to site 0. Here, the index $n$ runs over all $N$ sites in the real space. The operators $b^\dagger_n$ and $b_n$ are the bosonic creation and annihilation operators, respectively, for the harmonic bath at site $n$. The parameter $J$ represents the hopping amplitude between nearest-neighbor sites, while $c^\dagger_n$ and $c_n$ denote the fermionic creation and annihilation operators at site $n$. The coupling strength between the electronic carrier at site $n$ and the harmonic mode at the same site is given by $g$. The harmonic baths are assumed to be dispersionless, with a constant frequency $\omega_0$.
Taking the lattice constant as 1, the Hamiltonian in Eq.~\eqref{eqn:realspace_Holstein_Hamiltonian} can be equivalently written in momentum space as 
\begin{align}
    H
     &=\sum_{k} \left[ -2J(\cos{k}) \hat{c}^\dagger_k\hat{c}_k + \sum_q \frac{g}{\sqrt{N}}\hat{c}^\dagger_{k+q}\hat{c}_k (\hat{b}^\dagger_{-q} + \hat{b}_{q}) + \omega_0 \hat{b}^\dagger_k \hat{b}_k \right],
\end{align}
where $k$ and $q$ are momenta of the electrons and phonons.

In the limit of a single electronic particle excitation, the real-space Hamiltonian in Eq.~\eqref{eqn:realspace_Holstein_Hamiltonian} becomes
\begin{align}
    H&\equiv\sum_{n=0}^{N-1} -J\left(\ketbra{n+1}{n} + h.c. \right)+ g\ketbra{n}{n} (\hat{b}^\dagger_n + \hat{b}_n)+ \omega_0 \hat{b}^\dagger_n \hat{b}_n,
\end{align}
where $\ket{n}=\ket{\emptyset}_0 \otimes \cdots \otimes \ket{1}_n \otimes \cdots \otimes \ket{\emptyset}_{N-1}$ means an occupation of one electron on site $n$ while the other sites remain empty.
The thermodynamic limit may be approximated for large $N$. Empirically, $N \approx 16$ is sufficient to limit all finite-size effects for the parameters used here.~\cite{Mitri2023} The study of small $N$ cases, namely ``Holstein molecules,'' is interesting in its own right, as finite-size effects could represent physical manifestations of spectral behavior in small molecules. In this work, both limits are discussed.
\subsubsection{\textit{Ab Initio} Hamiltonian for LiF}
LiF is a polar ionic crystal with a large band gap (14.2 eV experimental optical gap). We consider the case of a single hole in the valence bands close to the band maximum. It is known that holes form small, strongly coupled polarons in this setting. The LiF hole Hamiltonian is given by
\begin{align}   H&=\sum_{nk}\epsilon_{nk}\hat{c}^\dagger_{nk}\hat{c}_{nk}+\sum_{\nu k}\omega_{\nu k}\hat{b}^\dagger_{\nu k}\hat{b}_{\nu k} \nonumber \\
&+\sum_{\substack{mn\nu\\kq}} g_{mn\nu kq} \hat{c}^\dagger_{mk+q}\hat{c}_{nk}(\hat{b}^\dagger_{\nu, -q}+\hat{b}_{\nu q}), \label{eq:H_lif}
\end{align}
where $m,~n$ label the hole bands, $\nu$ labels the phonon bands, and $k,~q$ denote the momenta of the hole and phonon degrees of freedom. We consider 3 hole bands and the coupling to longitudinal optical (LO) phonon modes in our MQC and variational Monte Carlo (VMC) implementations. Details of the VMC method are described in the supplementary materials (SM). The Hamiltonian parameters were obtained from density functional perturbation theory calculations using Quantum Espresso with the EPW package (see Ref. \citenum{robinson2025ab} for details where the same parameterization was used.). Due to the polar nature of this system, the hole coupled to the LO mode is the largest contributor to polaron formation, and this coupling is essentially of the Fr\"{o}hlich type.
\subsection{Spectral function}
We define the correlation function $C_k(t)$ with $\hbar=1$ as
\begin{align}
    C_k(t)=\frac{\operatorname{Tr} \left[ e^{-\beta H} e^{iHt}\hat{c}_k e^{-iHt}\hat{c}_k^\dagger \right]}{\operatorname{Tr} [ e^{-\beta H}]}. \label{eqn:ck}
\end{align}
In the limit of a single particle in a band, this becomes
\begin{align}
    C_k(t)=\frac{\operatorname{Tr} \left[ e^{-\beta H_B} e^{iHt}\hat{c}_k e^{-iHt}\hat{c}_k^\dagger \right]}{\operatorname{Tr} [ e^{-\beta H_B}]}, \label{eqn:ck_single}
\end{align}
where $H_B$ is the purely bath term in the Hamiltonian $H$. The derivation of Eq.~\eqref{eqn:ck_single} from Eq.~\eqref{eqn:ck} can be found in the SM.
The correlation function is related to the finite temperature retarded one-particle Green's function $G(k,t)$ as
\begin{align}
    G(k,t)= -i\theta(t) C_k(t),
\end{align}
where $\theta(t)$ is the Heaviside step function.
Upon Fourier transforming the Green's function, we obtain the spectral function
\begin{align}
    A(k,\omega)= -\frac{1}{\pi}\Im \left[ \int_{0}^\infty dt e^{i\omega t} e^{-\gamma t} G(k,t) \right].
\end{align}
Here, $\gamma$ is a broadening parameter. Unless otherwise specified, we set $\gamma=0.05$ in this work. 
In this paper, we focus on calculating the spectral function $A(k,\omega)$ for different models using various exact and mixed quantum-classical methods. These methods are described below.
\subsection{Mean Field Ehrenfest Dynamics} \label{sec:theorymfe}
We start with a generic Hamiltonian of a system interacting with multiple baths,
\begin{align}
    H&=\sum_{ij} H_{S,ij}(\hat{q}_i,\hat{p}_j)+\sum_{ij} H_{B,ij}(\hat{Q}_i,\hat{P}_j) \nonumber\\
    &+\sum_{ii'jj'}H_{SB,ii'jj'}(\hat{q}_i,\hat{p}_{i'},\hat{Q}_j,\hat{P}_{j'}),
\end{align}
where $\hat{P}_i$ and $\hat{p}_i$ are the mass-normalized momentum operators of the $i^{\text{th}}$ bath and system degree of freedom, respectively, while $\hat{Q}_i$ and $\hat{q}_i$ are their corresponding mass-normalized position operators.
We next replace the bath operators with bath variables, whose dynamics is classical, to define the mixed quantum-classical Hamiltonian,
\begin{align}
    H_{\text{MQC}}&=\sum_{ij} H_{S,ij}(\hat{q}_i,\hat{p}_j)+\sum_{ij}H_{B,ij}(Q_i,P_j)\nonumber\\
    &+\sum_{ii'jj'}H_{SB,ii'jj'}(\hat{q}_i,\hat{p}_{i'},Q_j,P_{j'}). \label{mixedH}
\end{align}
Denoting
\begin{align}
H_{S}&=\sum_{ij} H_{S,ij}(\hat{q}_i,\hat{p}_j),\\
H_{SB}&=\sum_{ii'jj'}H_{SB,ii'jj'}(\hat{q}_i,\hat{p}_{i'},Q_j,P_{j'}),\\
H_{B}&=\sum_{ij} H_{B,ij}(Q_i,P_j),
\end{align}
the classical equations of motion for the $i^\text{th}$ bath variables are
\begin{align}
    \dot{P_i}
    &= - \bra{\psi} \nabla_{Q_i} H_{SB} \ket{\psi} - \frac{\partial H_B}{\partial Q_i}, \label{P-evol}
\\
\dot{Q_i}
        &= \bra{\psi} \nabla_{P_i} H_{SB} \ket{\psi} + \frac{\partial H_B}{\partial P_i}, \label{Q-evol}
\end{align}
and the system is propagated with a Schrodinger equation of the form
\begin{align}
    i\frac{d}{dt}\ket{\psi(t)} = \left(H_{S}+H_{SB}(\vec{Q},\vec{P})\right)\ket{\psi(t)}. \label{eqn_main:schrodinger}
\end{align}
The change in the system's wavefunction directly influences the force acting on the bath positions and momenta, while the evolution of the bath variables affects the Hamiltonian that governs the propagation of the system's wavefunction. The system and the bath are propagated alternately at each time step according to Eq.~\eqref{P-evol}, Eq.~\eqref{Q-evol}, and Eq.~\eqref{eqn_main:schrodinger}. 
In this work, the initial bath positions and momenta are sampled from the Wigner distribution $\rho_W (\vec{P}, \vec{Q}, \beta)$ of a harmonic bath at an inverse temperature $\beta$.
For the approach we refer to in the following as the frozen bath method, the bath variables $\vec{P}$ and $\vec{Q}$ remain fixed at their initially sampled values throughout the propagation.~\cite{Zacharias2015} Meanwhile, in the MFE method without back-reaction, $\vec{P}$ and $\vec{Q}$ evolve with the system-bath coupling term $H_{SB}$ set to zero.

To compute $C_k(t)$ within MFE, we initialize the wavefunction in state $\ket{k}$ with a well-defined wave vector $k$. $C_k(t)$ is equivalent to the coefficient of the $\ket{k}$ state at time $t$,
\begin{align}
    C_k(t)=\int d\vec{P}_0 d\vec{Q}_0 \rho_W(\vec{P}_0,\vec{Q}_0,\beta) \langle k | e^{-i(H_S+H_{SB})t} | k \rangle. \label{eqn_main:mfecorr}
\end{align}
The reader is referred to the SM for a derivation of Eq.~\eqref{eqn_main:mfecorr}.

\subsection{The Multi-State Mapping Approach to Surface Hopping} \label{sec:theorymash}
In this section, we begin by reviewing the Multi-State Mapping Approach to Surface Hopping~\cite{Mannouch2023,Runeson2023,Runeson2024,Richardson2025} (MASH), particularly the formalism developed by Runeson \textit{et al.}~\cite{Runeson2023,Runeson2024} We then introduce two distinct ways to employ MASH to compute $A(k,\omega)$. Although these methods approach the problem from different angles, we find that they provide identical spectral functions numerically.
\subsubsection{Overview of the Algorithm}
As is typical of mixed quantum-classical methods, MASH propagates the electronic wavefunction quantum mechanically, as in Eq.~\eqref{eqn_main:schrodinger}, while treating the bath variables classically. However, MASH differs from MFE in two important ways. First, the equations of motion in MASH for bath variables take the form of
\begin{align}
    \dot{P_i}
    &= - \bra{\alpha} \nabla_{Q_i} H_{SB} \ket{\alpha} - \frac{\partial H_B}{\partial Q_i}, \label{eqn_main:p-evol}\\
\dot{Q_i}&= \bra{\alpha} \nabla_{P_i} H_{SB} \ket{\alpha} + \frac{\partial H_B}{\partial P_i},\label{eqn_main:q-evol}
\end{align}
where $\ket{\alpha}$ is the current active adiabatic state. Unlike the average force in MFE, the force in MASH arises from one adiabatic state at a time.
Although MASH is derived from Tully's Fewest Switches Surface Hopping~\cite{Tully1990}, where hops are stochastic, it is deterministic. The state $|\alpha\rangle$ is selected based on the adiabatic surface with the highest population, except when energy conservation prohibits such a hop.
Hops are carried out by switching the active surface and adjusting the momentum to conserve the total energy. The element of the vector along which the momentum is rescaled is
\begin{align}
    v_j=\sum_{a'}\Re[c^*_{a'}(d^j_{a'a}c_a-d^j_{a'b}c_b)],
\end{align}
where $c_i$ are the coefficients of the wavefunction in the adiabatic basis, and
\begin{align}
    d^j_{a'a}=\langle a'| \nabla_j | a \rangle.
\end{align}
If the kinetic energy is insufficient to overcome the potential difference between the old and new active surfaces, the momentum's direction will be reversed in the opposite direction of the above vector without any rescaling of the magnitude, and the hop is canceled.

Second, in contrast to MFE, MASH involves the stochastic sampling of the initial bath variables as well as the initial coefficients of the system's wavefunction. We refer to Ref.~\onlinecite{Runeson2024} for the specifics on the different sampling schemes for the wavefunction. In practice, we have found essentially no differences in the results between these schemes, and thus have utilized focused initial conditions to sample the wavefunction coefficients. 
In particular, we sample a wavefunction focused on state $\ket{k}$ from the following density distribution,
\begin{align}
    \rho_k(c)=\delta \left( |c_k|^2 - \frac{1+\beta_N}{\alpha_N} \right) \prod_{i \neq k} \delta \left( |c_i|^2 - \frac{\beta_N}{\alpha_N} \right), \label{eq:MASH_sampling}
\end{align}
where
\begin{align}
    \alpha_N&=\frac{N-1}{\sum_{n=1}^N \frac{1}{n}-1},\\
    \beta_N&=\frac{\alpha_N-1}{N},
\end{align}
and $N$ is the size of the system.
The phases of the coefficients are sampled uniformly over $[0,2\pi)$. 

\subsubsection{Estimator of the Electronic Wavefunction Coefficients}
One way to compute the correlation function $C_k(t)$ is to use an estimator for the electronic wavefunction coefficients. In this section, we give a general derivation of the estimator for the electronic wavefunction coefficients, while giving a specific example on how it is applied to calculate $C_k(t)$.
For a given initial arbitrary wavefunction $\ket{\psi_0}$, we consider a superposition, normalized state
\begin{align}
    \ket{\psi(0)}=m_\phi \ket{\phi} + m_{\psi_0} \ket{\psi_0}, \label{eqn:MASH_initial_extended_equation}
\end{align}
where $|m_\phi|^2+|m_{\psi_0}|^2=1$, and $\ket{\phi}$ is an auxiliary state uncoupled to the original Hilbert space of the system in the Hamiltonian, meaning that the Hamiltonian does not act on the $\ket{\phi}$ state. In the limit of $m_\phi \rightarrow 0$ and $m_{\psi_0} \rightarrow 1$, we recover the original state $\ket{\psi_0}$. In our specific context of calculating $C_k(t)$ according to Eq.~\eqref{eqn_main:mfecorr}, we set $\ket{\psi_0}=|k\rangle$, and sample the initial state as follows,
\begin{align}
    \rho(c)&=\delta \left( |c_{\tilde{k}}|^2 - \frac{1+\beta_{N+1}}{\alpha_{N+1}} \right) \delta \left( |c_{\tilde{\phi}}|^2 - \frac{\beta_{N+1}}{\alpha_{N+1}} \right) \nonumber \\
    &\prod_{i \neq k} \delta \left( |c_i|^2 - \frac{\beta_{N+1}}{\alpha_{N+1}} \right),\label{eq:MASH_Ck_sampling}
\end{align}
where
\begin{align}
    |\tilde{k}\rangle&=\ket{\psi(0)}=m_\phi \ket{\phi} + m_{\psi_0} \ket{k},\\
    |\tilde{\phi}\rangle&=m_{\psi_0} \ket{\phi} - m_{\phi} \ket{k},
\end{align}
The index $i$ runs through all the states in the momentum basis that are orthonormal to state $|k\rangle$. It is important to note that the magnitudes of the coefficients \textit{include} the auxiliary state. We take $m_\phi$ and $m_{\psi_0}$ to be real numbers, with $m_\phi^2+m_{\psi_0}^2=1$, so that the tilde basis is orthonormal.

Returning to our general formulation of the electronic coefficient estimator, from Ref.~\onlinecite{Runeson2023}, we have the following estimators for the diagonal and off-diagonal elements of a system observable,
\begin{align}
    \Phi_{nn}&=\frac{1}{N+1}+\alpha_{N+1} \left( |c_n|^2-\frac{1}{N+1}\right), \label{eq:MASH_pop_extended}\\
    \Phi_{nm}&=\alpha_{N+1} c^*_n c_m, \label{eq:MASH_coherence_extended}
\end{align}
where $\Phi_{nn}$ and $\Phi_{nm}$ estimate the elements $\ketbra{n}{n}$ and $\ketbra{n}{m}$ respectively, and $c_n$ is the coefficient of state $\ket{n}$. The use of $N+1$ here instead of $N$ in the original formulation in Ref.~\onlinecite{Runeson2023} explicitly emphasizes the inclusion of the auxiliary state in our formulation.
We consider an element of the density matrix constructed from the wavefunction in Eq.~\eqref{eqn:MASH_initial_extended_equation},
\begin{align}
    \rho_{i\phi} (t)&=\expecth{i}{e^{-iHt}}{\psi(0)}\expecth{\psi(0)}{e^{iHt}}{\phi}\\
    &=m_{\psi_0} m_\phi\bra{i}e^{-i H t}\ket{\psi_0}\\
    &=m_{\psi_0} m_\phi c_i(t).
\end{align}
$\rho_{i\phi} (t)$ is estimated by the population or coherence estimator $\Phi_{\phi i} (t)$ as expressed in Eq.~\ref{eq:MASH_pop_extended} and Eq.~\ref{eq:MASH_coherence_extended}.
We obtain
\begin{align}
\expect{c_i(t)}= \frac{\left\langle \rho_{i\phi} (t) \right \rangle_\text{MQC}}{m_{\psi_0} m_\phi} = \frac{\left\langle \Phi_{\phi i} (t) \right\rangle_\text{MASH}}{m_{\psi_0} m_\phi} = \left\langle \frac{c_i(t)}{c_{\psi_0}(0)} \right\rangle_\text{MASH}. \label{eqn_main:ci}
\end{align}
The derivation of Eq.~\eqref{eqn_main:ci} can be found in the SM. We can now compute the estimator for $\expecth{i}{e^{-iHt}}{\psi_0}$ as
\begin{align}
    W_{i\psi_0} (t) = \frac{c_i (t)}{c_{\psi_0}(0)}.
\end{align}
Since this estimator is based on the population or coherence estimator~\cite{Runeson2023}, it is basis-equivariant.

Although $m_\phi$ and $m_{\psi_0}$ do not explicitly appear in the final expression for the electronic coefficient estimator, they might influence it indirectly. For example, in the case of computing the correlation function $C_k(t)$, after the initial sampling by Eq.~\eqref{eq:MASH_Ck_sampling}, we rotate this basis to the real space basis, and MASH propagates the system in this basis. Note that real space states $|n\rangle$ are Fourier transforms of physical momentum states $|k\rangle$ only, excluding the auxiliary state $|\phi\rangle$. The initial wavefunction now becomes
\begin{align}
|\Psi(0)\rangle 
&=\left(m_\phi\sqrt{\frac{1+\beta_{N+1}}{ \alpha_{N+1}}} e^{i \theta_{\tilde{k}}}-m_k\sqrt{\frac{\beta_{N+1}}{ \alpha_{N+1}}} e^{i \theta_{\tilde{\phi}}}\right)|\phi\rangle \nonumber \\
& +\sum_n m_k\sqrt{\frac{1+\beta_{N+1}}{ \alpha_{N+1} N}} e^{i \theta_{\tilde{k}}} e^{-i k n}|n\rangle\nonumber\\
&+\sum_n m_\phi\sqrt{\frac{\beta_{N+1}}{ \alpha_{N+1} N}} e^{i \theta_{\tilde{\phi}}} e^{-i k n}|n\rangle\nonumber\\ 
&+\sum_n\sum_{q} \sqrt{\frac{\beta_{N+1}}{\alpha_{N+1} N}} e^{i \theta_{q} }e^{-i q n}|n\rangle.
\end{align}
One can observe that due to the interactions of $m_\phi$, $m_k$ and the initial sampled phases, such as $\theta_{\tilde{\phi}}$ and $\theta_{\tilde{k}}$, the relative populations of the states $|n\rangle$ might be different in each trajectory, influencing the selection of the adiabatic surface for the MASH force. This happens even though the system never interacts with the state $|\phi\rangle$ in our MASH algorithm. We also note that in our MASH algorithm, the system never hops to this uncoupled surface even if it carries the highest population.
To remain as close as possible to the original physical system, one without the uncoupled state, we can initialize the system in a superposition with minimal mixing with the $\ket{\phi}$ state, where $m_\phi \rightarrow 0$ and $m_{\psi_0} \rightarrow 1$. However, in practice, the specific choice of $m_\phi$ and $m_{\psi_0}$ has little impact on the estimator for the electronic coefficient. In this work, we set $m_\phi=\frac{1}{\sqrt{2}}$ and have found no qualitative difference compared to the spectra obtained at $m_\phi=10^{-5}$. Representative examples are provided in the SM.

From the derivation of the correlation function in MFE (Eq.~\eqref{eqn_main:mfecorr}), we see that the estimator for the correlation function in MASH is really the estimator for $\expecth{k}{e^{-i(H_S+H_{SB})t}}{k}$. The expression for the correlation function in MASH is therefore
\begin{align}
    C_k(t)&=\int d\vec{P}_0 d\vec{Q}_0 \rho_W(\vec{P}_0,\vec{Q}_0, \beta) \int d\vec{c}  \rho_k(\vec{c}) W_{kk}(t)\nonumber\\
    &=\int d\vec{P}_0 d\vec{Q}_0 \rho_W(\vec{P}_0,\vec{Q}_0, \beta) \int d\vec{c}  \rho_k(\vec{c}) \frac{c_k(t)}{c_k(0)}.
\end{align}

\subsubsection{Computation of the correlation function via the pure-state decomposition}
Another way to compute the correlation function $C_k(t)$ is via the pure-state decomposition. Our numerical calculations show that the spectra calculated via the pure-state decomposition and the wavefunction coefficient estimator method give indistinguishable results in both MFE and MASH (see SM).
The creation and annihilation operators $\hat{c}^\dagger_k$ and $\hat{c}_k$ can be rewritten as $\ketbra{k}{\emptyset}$ and $\ketbra{\emptyset}{k}$ respectively in the single-particle limit.
Following work by Lieberherr \textit{et al.}~\cite{manounpub} as well as previous studies~\cite{Atsango2023,MontoyaCastillo2016}, we decompose these operators into linear combinations of pure-state density matrices as follows,
\begin{align}
    \ketbra{a}{b}=\sum_{j=0}^3w_j\ketbra{j}{j},
\end{align}
where
\begin{align}
    \ket{j}&=\frac{1}{\sqrt{2}}\left(\frac{\ket{a}}{\sqrt{\braket{a}{a}}}+e^{ij\frac{\pi}{2}}\frac{\ket{b}}{\sqrt{\braket{b}{b}}}\right),\\
    w_j&=\frac{\sqrt{\braket{a}{a}\braket{b}{b}}}{2}e^{ij\frac{\pi}{2}}.
\end{align}
In our case of the correlation function, we define
\begin{alignat}{4}
    w_0 = \frac{1}{2},     &\quad 
    w_1 = \frac{i}{2},      &\quad 
    w_2 = \frac{-1}{2},    &\quad 
    w_3 = \frac{-i}{2},
\end{alignat}
and
\begin{alignat}{2}
    \ket{0}=\frac{1}{\sqrt{2}}(\ket{\emptyset}+\ket{k}), &\quad
    \ket{1}=\frac{1}{\sqrt{2}}(\ket{\emptyset}+i\ket{k}),&\quad\\
    \ket{2}=\frac{1}{\sqrt{2}}(\ket{\emptyset}-\ket{k}),&\quad
    \ket{3}=\frac{1}{\sqrt{2}}(\ket{\emptyset}-i\ket{k}).
\end{alignat}
The correlation function is computed as
\begin{align}
    C_k(t)&=\frac{\operatorname{Tr} \left[ e^{-\beta H_B} e^{iHt}\ketbra{\emptyset}{k} e^{-iHt}\ketbra{k}{\emptyset} \right]}{\operatorname{Tr}_B [ e^{-\beta H_B}]}\\
    &=\frac{\operatorname{Tr}_B \left[ e^{-\beta H_B} \sum_{j=0}^3 w_jw_0\braket{j}{0(t)}\braket{0(t)}{j}\right]}{\operatorname{Tr}_B [ e^{-\beta H_B}]} \nonumber\\
    &-\frac{\operatorname{Tr}_B \left[ e^{-\beta H_B} \sum_{j=0}^3 w_jw_0\braket{j}{2(t)}\braket{2(t)}{j}\right]}{\operatorname{Tr}_B [ e^{-\beta H_B}]}\nonumber\\
    &+\frac{\operatorname{Tr}_B \left[ e^{-\beta H_B} \sum_{j=0}^3 w_jw_0i\braket{j}{+i(t)}\braket{+i(t)}{j}\right]}{\operatorname{Tr}_B [ e^{-\beta H_B}]}\nonumber\\
    &-\frac{\operatorname{Tr}_B \left[ e^{-\beta H_B} \sum_{j=0}^3 w_jw_0i\braket{j}{-i(t)}\braket{-i(t)}{j}\right]}{\operatorname{Tr}_B [ e^{-\beta H_B}]}\label{eqn:pure-state-corr-general}\\
    &=\frac{\operatorname{Tr}_B \left[ e^{-\beta H_B} \sum_{j=0}^3 w_jw_0 \alpha_{N+1}\left(|c^{0}_j|^2 - |c^{2}_j|^2 \right)  \right]}{\operatorname{Tr}_B [ e^{-\beta H_B}]}\nonumber\\
    &+\frac{\operatorname{Tr}_B \left[ e^{-\beta H_B} \sum_{j=0}^3 w_jw_0 \alpha_{N+1}i\left(|c^{+i}_j|^2-|c^{-i}_j|^2 \right) \right]}{\operatorname{Tr}_B [ e^{-\beta H_B}]}.\label{eqn:pure-state-corr}
\end{align}
Here, $c^n_j$ represents the coefficient of state $\ket{j}$ when the initial state is $\ket{n}$, and $\ket{\pm i}=\frac{1}{\sqrt{2}}(\ket{0}\pm i\ket{2})$.
To sample the initial states $|j(0)\rangle$, we follow Eq.~\eqref{eq:MASH_Ck_sampling} as in the estimator method. The uncoupled state $|\phi\rangle$ is excluded from the hopping manifold and does not interact with the remaining states during the propagation.
We arrive at Eq.~\eqref{eqn:pure-state-corr} by taking the standard population estimator in Ref.~\onlinecite{Runeson2023} to express Eq.~\eqref{eqn:pure-state-corr-general}. For the derivation of Eq.~\eqref{eqn:pure-state-corr-general}, the reader is referred to the SM.

\subsection{Exact Analytical Solutions in The Regime of Zero Hopping Amplitude and Zero Temperature} \label{sec:theoryanalytical_J=0T=0}
Analytical expressions for the spectral function in MFE, MASH, and in the exact quantum case can be derived for the one-dimensional Holstein model in the simplified limit where the hopping amplitude $J$ is set to 0. For simplicity, we set $T=0$ as well. In this regime, the sites are completely decoupled, resulting in a spectral function that is independent of variation in $k$. For the exact quantum case, it is simple to show that the spectral function is
\begin{align}
    A(\omega)= e^{-g^2} \sum_{n=0}^\infty \frac{g^{2n}}{n!} \delta(\omega+g^2\omega_0-\omega_0 n), \label{spec_exact}
\end{align}
and is independent of $N$.
Despite the fact that the sites are completely decoupled for $J=0$, MFE and MASH produce spectral functions that are dependent on $N$ in different ways. This dependency is an artifact of the methods themselves rather than a physically meaningful phenomenon and arises due to the way each method treats the initial superposition state, $\ket{k}=\frac{1}{\sqrt{N}}\sum_ne^{ikn}\ket{n}$. In particular, MFE yields exact results when the number of sites is 2 (as shown in the upper two rows of Fig.~\ref{fig:MFEvsMASH_J}). For arbitrary $N$, the MFE spectral function can be calculated analytically,
\begin{align}
    A(\omega)=  e^{-g^2} &\sum_{n=-\infty}^{\infty} \sum_{m=-\infty}^{\infty} I_m\left(g^2\right) J_{n-m}\left(-\frac{2}{N} g^2\right) \nonumber \\
    &\delta\left(\omega+2 \frac{g^2}{N} \omega_0+n \omega_0\right), \label{spec_mfe}
\end{align}
where $J_n$ and $I_n$ are the $n^{th}$ Bessel function and modified Bessel function of the first kind, respectively. For $N=2$, Eq.~\eqref{spec_mfe} reduces to Eq.~\eqref{spec_exact}.

The modified MASH approach also fails to reproduce the exact results in general. In the scenario where $g^2$ is an integer (as shown in the lower two rows of Fig.~\ref{fig:MFEvsMASH_J}), MASH produces peaks that align well with the exact ones. However, the weights of the peaks are inaccurate, and additional peaks emerge that are not present in the exact results. The spectral function produced by MASH can be analytically determined to be
\begin{alignat}{2}
    A(\omega) =\frac{1}{N} & e^{-g^2} \sum_{n=-\infty}^{\infty} \sum_{m=-\infty}^{\infty} && I_m\left(g^2\right) J_{n-m}\left(-2 g^2\right) \nonumber \\
    & &&\delta\left[\omega+2 g^2 \omega_0+\omega_0 n\right] \nonumber \\
    + & \frac{N-1}{N} e^{-g^2} \sum_{n=-\infty}^{\infty} && I_n\left(g^2\right) \delta[\omega+\omega_0n]. \label{spec_mash}
\end{alignat}
Note that both MFE and MASH may produce regions of negative spectral weight, and this is reproduced analytically for the $J=0$ case, a fact that implies the associated regions of negative spectral weight seen in Fig.~\ref{fig:MFEvsMASH_J} and Fig.~\ref{fig:MFEvsMASH_Tg} do not arise from numerical undersampling. Interestingly, these regions are more noticeable in MFE than in MASH in a few regimes.
For derivations of the spectral functions in Eq.~\eqref{spec_exact}, Eq.~\eqref{spec_mfe}, and Eq.~\eqref{spec_mash}, the reader is referred to the SM.

\section{Results and Discussion}\label{sec:results}
In this section, we begin by comparing MFE and MASH, focusing on the effect of changing the hopping parameter $J$, the coupling strength $g$, temperature $T$, and the lattice size $N$ on the performance of the two methods in the one-dimensional Holstein model. For MASH, both $m_k$ and $m_\phi$ are set to $\frac{1}{\sqrt{2}}$. Next, we compare the spectral functions computed by MFE, frozen MFE, MFE without the back-reaction force, and the cumulant expansion (CE) method for small and large lattice sizes. Finally, we calculate the spectral function for an \textit{ab initio} model of a hole in the upper valence bands of LiF with MFE and compare the result to previous cumulant calculations as well as a near-exact benchmark result calculated via variational quantum Monte Carlo. We use 10,000 trajectories for all MFE variants applied to the Holstein model, 30,000 trajectories for MASH, and 200 trajectories for MFE applied to LiF to achieve convergence.

\begin{figure*}[t]
    \centering
    \includegraphics[width=\textwidth]{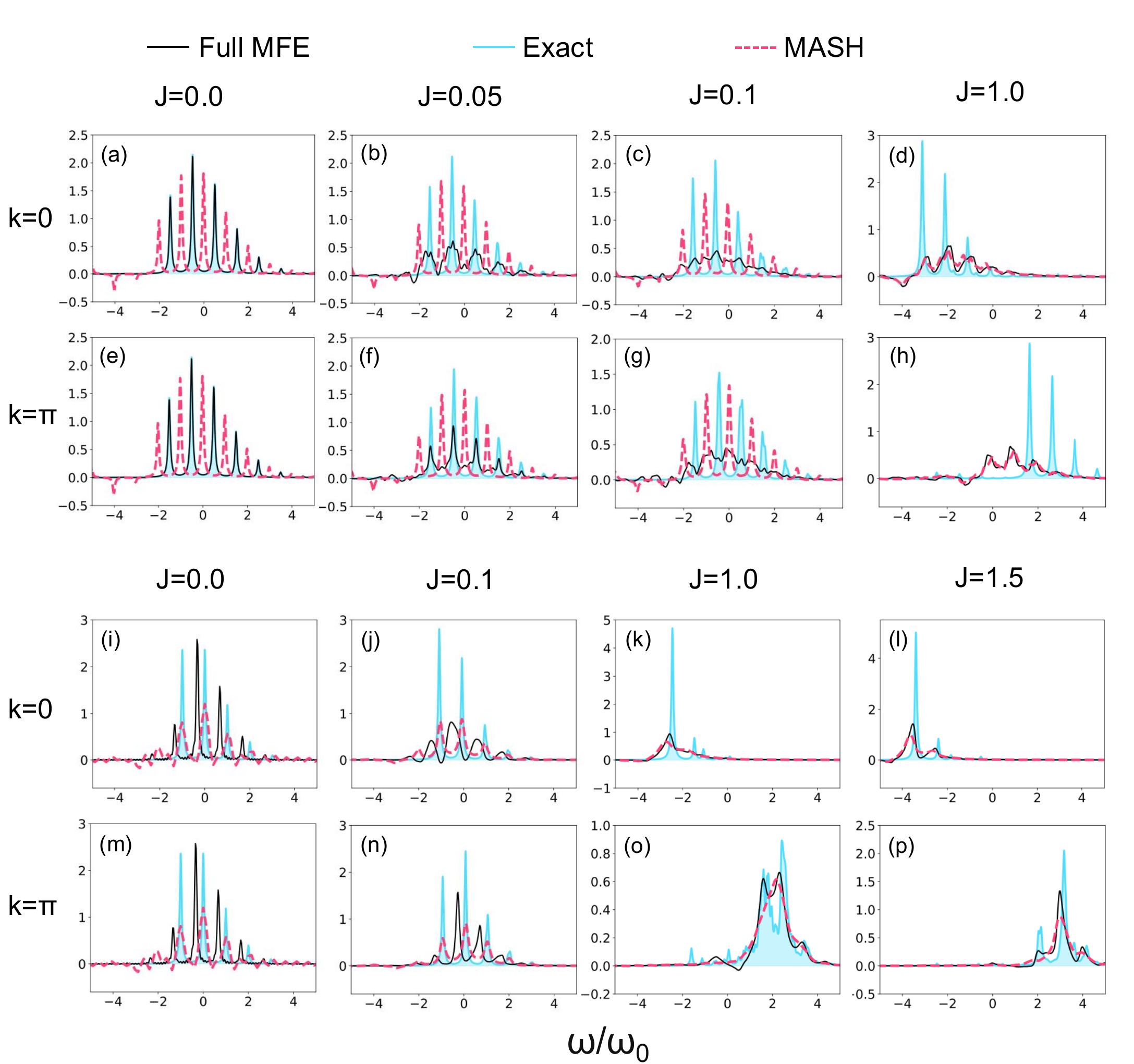}
    \caption{Spectral functions computed by MFE, MASH, and exact diagonalization, plotted across a range of hopping parameters $J$ and $k$-points at zero temperature. The upper two rows ((\textbf{a})--(\textbf{h})) depict results for the 2-site Holstein model with a bath frequency $\omega_0=0.91$ and coupling strength $g=1.23$, while the the lower two rows ((\textbf{i})--(\textbf{p})) show results for the 6-site Holstein model with $\omega_0=g=1.0$. For each regime, the top row presents results for $k=0$, and the bottom row present results for $k=\pi$.}
    \label{fig:MFEvsMASH_J}
\end{figure*}

\begin{figure*}[t]
    \centering
    \includegraphics[width=\textwidth]{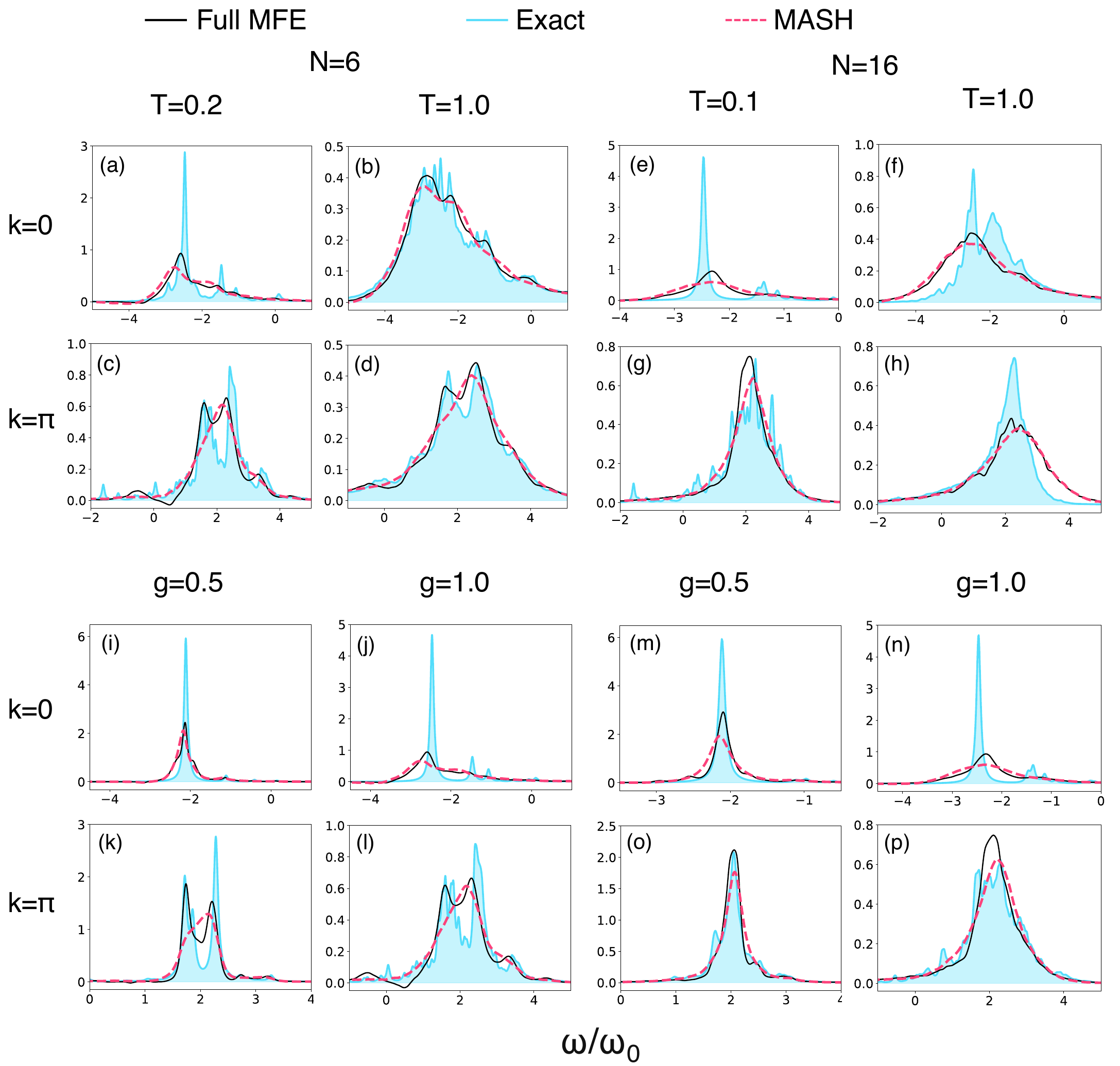}
    \caption{Spectral functions computed using MFE, MASH, and exact diagonalization. The upper two rows illustrate results for the Holstein model with the parameters $J=g=\omega_0=1.0$ across different temperatures. (\textbf{a})--(\textbf{d}) show results for a 6-site lattice, while (\textbf{e})--(\textbf{h}) show results for a 16-site lattice. For each lattice size, the left column represents low temperature while the right column represents high temperature. In the lower two rows, the model is evaluated at zero temperature with $J=\omega_0=1.0$ across different coupling strengths $g$, where (\textbf{i})--(\textbf{l}) correspond to 6 sites and (\textbf{m})--(\textbf{p}) to 16 sites. For each lattice size, the left column corresponds to $g=0.5$ while the right column corresponds to $g=1.0$.}
    \label{fig:MFEvsMASH_Tg}
\end{figure*}

Starting from the limit of $J=0$ where exact answers are known, we increase $J$ to assess the comparative performance of MFE and MASH. Fig.~\ref{fig:MFEvsMASH_J} shows the spectral functions obtained using exact diagonalization, MFE, and MASH in different regimes of small periodic Holstein systems at zero temperature. The upper two rows, Fig.~\ref{fig:MFEvsMASH_J}(a--h), present results for the 2-site Holstein model with the coupling strength $g=1.23$ and the bath mode frequency $\omega_0=0.91$. These values are chosen so as to avoid spurious alignment with the exact peak structure. MFE gives exact results at $J=0$ for $N=2$ including in this regime. 
In contrast, Fig.~\ref{fig:MFEvsMASH_J}(i--p) show results for 6-site Holstein system with $g=1.0$ and $\omega_0=1.0$, where MASH gives peaks aligned with the exact solutions.

Fig.~\ref{fig:MFEvsMASH_J}(a--c,e--g) demonstrate that in the two-site Holstein model at very small $J$ values (up to 0.1), MFE is more accurate than MASH with respect to peak positions. In this range of $J$, MFE retains its ability to capture peak positions although the sharpness of the peaks decreases noticeably with creasing $J$. MASH, on the other hand, fares worse when $g^2$ is not an integer, producing misaligned and additional spurious peaks. As $J$ increases slightly from zero, MFE peaks begin to broaden, especially at $k=0$, as shown in Fig.~\ref{fig:MFEvsMASH_J}(b,c). In contrast, at $k=\pi$, the peaks remain relatively sharp initially (Fig.~\ref{fig:MFEvsMASH_J}f) but quickly broaden and shift as $J$ grows (Fig.~\ref{fig:MFEvsMASH_J}g). When $J$ reaches the standard value of 1.0, the performance of both MFE and MASH deteriorates significantly at $k=\pi$, as shown in Fig.~\ref{fig:MFEvsMASH_J}h. Interestingly, both methods produce very similar spectral functions here, a behavior also observed in the lower two rows when $J$ is sizable.

In the lower two rows, we present spectral functions for the 6-site Holstein model with the coupling strength $g=1.0$ and bath mode frequency $\omega_0=1.0$. In this regime, MFE does not generate peaks that align with the exact results at $J=0$, whereas MASH does. As $J$ increases to 0.1, MASH retains the correct peak locations, while MFE shows deviations from the exact positions. Although peak broadening becomes apparent in MFE, this effect is less pronounced in MASH, as shown in Fig.~\ref{fig:MFEvsMASH_J}(i,j,m,n). 
As $J$ increases beyond 1.0, MFE generally outperforms MASH. For example, at $J=1.0$ and $k=0$ (Fig.~\ref{fig:MFEvsMASH_J}k), MFE produces a sharper spectral function with the major peak closer to the exact result than that of MASH. At $k=\pi$ (Fig.~\ref{fig:MFEvsMASH_J}o), MFE successfully captures the split peaks, which MASH fails to resolve. This trend continues in the large hopping regime ($J=1.5$). At $k=0$ (Fig.~\ref{fig:MFEvsMASH_J}l), both methods capture the main peak and one large satellite peak, but the MFE spectral function is noticeably sharper. Finally, at $k=\pi$ (Fig.~\ref{fig:MFEvsMASH_J}p), MFE captures all three major peaks, whereas MASH only resolves the largest peak.

In conclusion, aside from cases deviating very slightly from $J=0$ in a regime where MASH will show an advantage, MFE is generally somewhat more accurate than MASH aside from the fact that small regions of negative spectral weight may be observed. With this established, we continue to assess other parameters (the number of sites $N$, the coupling strength $g$, and temperature $T$) in the Holstein model. Fig.~\ref{fig:MFEvsMASH_Tg} compares the spectral functions for 6 and 16 sites, keeping $J=1.0$ and varying coupling strength $g$ and temperature $T$. Note that $N=16$ is large enough to eliminate almost all finite-size effects. The upper two rows, Fig.~\ref{fig:MFEvsMASH_Tg}(a--h), present results across different temperatures, while the lower two rows, Fig.~\ref{fig:MFEvsMASH_Tg}(i--p), show results with varying coupling strengths.

Fig.~\ref{fig:MFEvsMASH_Tg}(a--d) present spectral functions at low and high temperatures at different \textit{k}-points in a chain of 6 sites. At $k=0$ and a low temperature (Fig.~\ref{fig:MFEvsMASH_Tg}a), MFE and MASH show comparable accuracy in representing the spectral function although the major peak in MASH is slightly more displaced from the exact result compared to MFE. At $k=0$ and a high temperature (Fig.~\ref{fig:MFEvsMASH_Tg}b), MFE and MASH produce very similar results, with MFE appearing to capture more spectral features. The performance of MFE is superior to that of that of MASH at $k=\pi$, as shown in Fig.~\ref{fig:MFEvsMASH_Tg}(c,d), again with the caveat that small negative regions appear at $k=\pi$ and low temperature. Here, MFE reproduces the split peaks while MASH is unable to resolve these features at both low and high temperatures.

Fig.~\ref{fig:MFEvsMASH_Tg}(e--h) present spectral functions for the same parameter regimes as in Fig.~\ref{fig:MFEvsMASH_Tg}(a--d), but with the number of sites increases from 6 to 16. At $k=0$ and low temperature, Fig.~\ref{fig:MFEvsMASH_Tg}e reveals a similar pattern to its 6-site counterpart in Fig.~\ref{fig:MFEvsMASH_Tg}a, with both MFE and MASH capturing the major peak, albeit with slight displacement. However, neither method captures the satellite peaks in the 16-site model, a notable contrast to the 6-site case. At high temperatures, as shown in Fig.~\ref{fig:MFEvsMASH_Tg}f, MASH and MFE perform less effectively, failing to accurately capture the split peaks of the 16-site model. Unlike the 6-site results shown in Fig.~\ref{fig:MFEvsMASH_Tg}b, the mixed quantum-classical spectral functions for the 16-site Holstein model at high temperature are slightly displaced. This displacement occurs not only at $k=0$ but also at $k=\pi$, as illustrated in Fig.~\ref{fig:MFEvsMASH_Tg}(f,h). Interestingly, this contrasts with the better accuracy observed at a low temperature, as shown in Fig.~\ref{fig:MFEvsMASH_Tg}g, making the high-temperature deviations somewhat surprising. MASH simulations initialized with a classical Boltzmann distribution in the high-temperature regimes of Fig.~\ref{fig:MFEvsMASH_Tg}(f,h) yield results consistent with those reported here, suggesting that nuclear overheating due to the use of the Wigner distribution for the initial bath variables is unlikely to account for the spectral broadening and displacement relative to the exact results.

\begin{figure}
    \centering
    \includegraphics[width=\columnwidth]{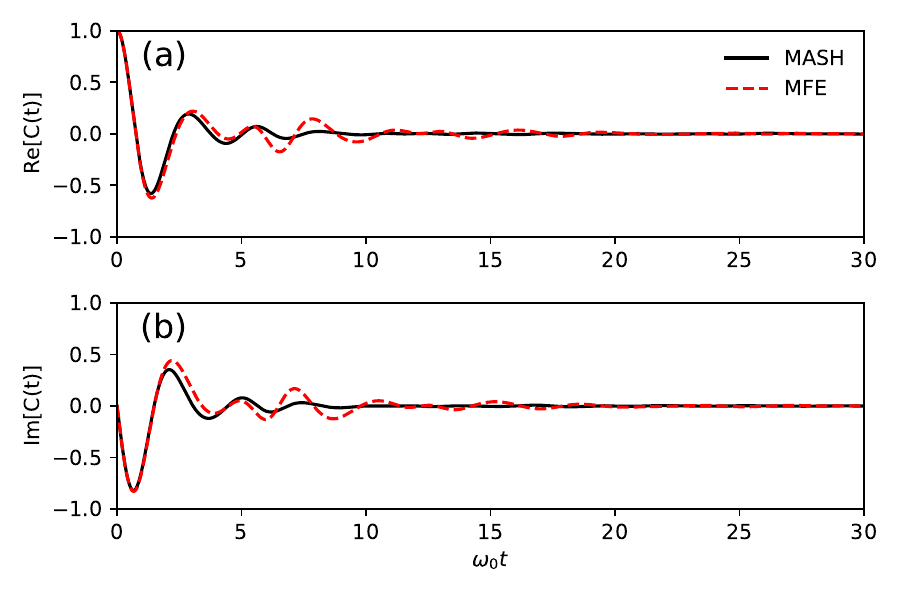}
    \caption{\textbf{(a)} Real and \textbf{(b)} imaginary parts of the correlation functions $C_k(t)$ computed using MFE and MASH for the 6-site Holstein model in the regime of $J=\omega_0=g=1.0$ at zero temperature and $k=\pi$.}
    \label{fig:Ct_k=pi}
\end{figure}

\begin{figure*}[t]
    \centering
    \includegraphics[width=\textwidth]{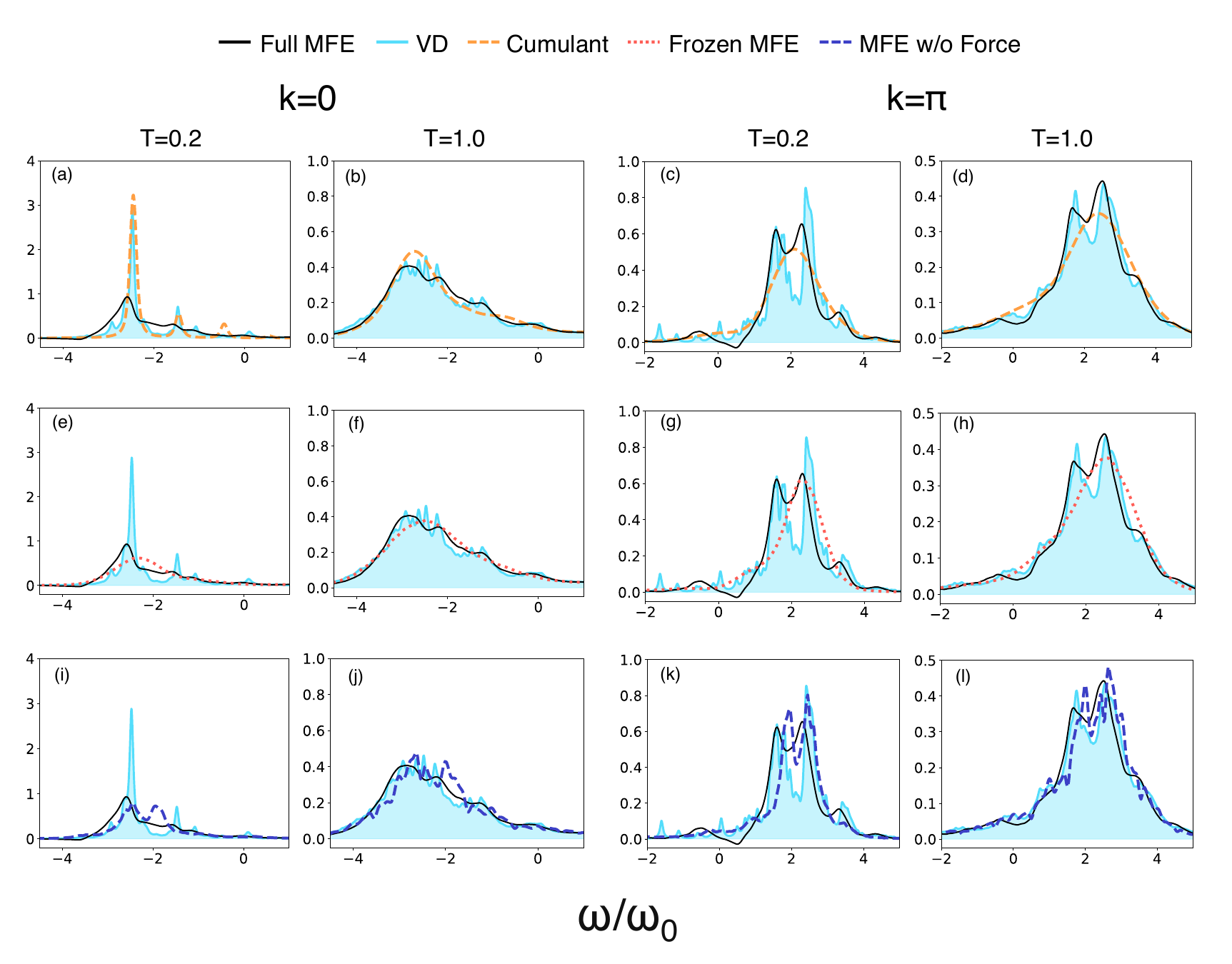}
    \caption{Spectral functions of a 6-site Holstein model with parameters $J = g = \omega_0 = 1.0$, evaluated at various temperatures and $k$-points. The left two columns correspond to $k=0$, while the right two columns correspond to $k=\pi$. For each $k$-point, the left column represents a temperature of 0.2, while the right column represents a temperature of 1.0. The first row ((\textbf{a})--(\textbf{d})) compares variational diagonalization (VD) results from Ref.~\onlinecite{Bona2019}, full MFE, and cumulant calculations from Ref.~\onlinecite{Robinson2022}. The second row ((\textbf{e})--(\textbf{h})) compares full MFE, frozen MFE, and VD. The third row ((\textbf{i})--(\textbf{l})) compares full MFE, MFE without the back-reaction force, and VD.}
    \label{fig:mfe_6sites}
\end{figure*}
In the lower two rows (Fig.~\ref{fig:MFEvsMASH_Tg}(i--p)), we explore the spectral functions at zero temperature, keeping $J=\omega_0=1.0$ while varying $g$. Overall, an increase in $g$ causes the spectral function to broaden and reveal more small satellite peaks. For the smaller ring of 6 sites, at $k=0$, MFE and MASH perform comparably, as shown in Fig.~\ref{fig:MFEvsMASH_Tg}(i,j). However, at $k=\pi$, MFE demonstrates clear superiority by successfully capturing the split peaks which MASH fails to resolve, as seen in Fig.~\ref{fig:MFEvsMASH_Tg}(k,l) as previously noted. For the larger 16-site system (Fig.~\ref{fig:MFEvsMASH_Tg}(m--p)), MFE and MASH produce comparable spectral functions.

In conclusion, MFE again appears to be slightly more accurate than MASH, particularly in the regimes of small system sizes and low temperatures, in terms of peak positions. However, MFE suffers more severely from the negative spectral weight issue here. 
From examining the analytical results for the $J=0$ case, we note that this problem persists in both MFE and MASH, indicating that it arises from the algorithms themselves rather than physical inaccuracies or numerical artifacts.
Regarding peak positions, MFE is known to suffer from the overcoherence problem~\cite{Subotnik2010}, which MASH resolves through its deterministic hopping between adiabatic states~\cite{Runeson2023}. 
In regimes where $N$ is small and the system is not at very high temperatures (as shown in the $N=6$ panels of Fig.~\ref{fig:MFEvsMASH_Tg}), the tendency of MFE to retain coherence might be advantageous in depicting a system that is slow to decohere on the short and intermediate timescales relevant for the computation of the spectral function. The time-domain correlation functions in these regimes exhibit much slower decoherence in MFE compared to MASH, which explains MFE's richer spectral structure, particularly at $k=\pi$. As illustrated in Fig.~\ref{fig:Ct_k=pi}, the MFE correlation function shows a revival in amplitude after the initial decay, while the MASH correlation function decays monotonically. This results in a split peak in MFE spectral function but only a broad peak in MASH, as evidenced in Fig.~\ref{fig:MFEvsMASH_J}(o).
However, as the number of sites or temperatures increases, the distinction between the performance of MFE and MASH significantly lessens. Higher temperatures allow thermal effects to dominate, rendering quantal aspects of the lattice motion less important while larger system sizes lead to a more continuous spectrum at low energies, giving rise to smoother spectral features. In addition, in the case of larger lattices, coherent beating between discrete electronic states~\cite{Robinson2022} is removed as the energy spacing is decreased, and the effective number of bath oscillators increases.  Both factors lead to the broader spectrum exhibited for larger systems.
In fact, for $N=16$, which is sufficient to reach convergence with respect to the number of lattice sites, the two methods yield largely similar spectra characterized by broadened peaks.

We next focus on a specific regime where MFE has successfully captured non-trivial spectral features as demonstrated in Fig.~\ref{fig:MFEvsMASH_Tg}, namely the Holstein model on a 6-site ring with $J=1.0$, $g=1.0$, and $\omega_0=1.0$. We examine the nature of phonon dynamics in MFE to see how various approximations might affect the performance of MFE.
Fig.~\ref{fig:mfe_6sites} compares the standard full MFE with the cumulant expansion (CE) data obtained by Robinson \textit{et al.}~\cite{Robinson2022}, as well as with frozen MFE, and MFE without the back-reaction force, using the exact variational diagonalization (VD) results taken from Bon\v{c}a \textit{et al.}~\cite{Bona2019} as a benchmark. The spectral function is evaluated at $T=0.2$ and $T=1.0$ for $k=0$ and $k=\pi$. 

The upper panels, Fig.~\ref{fig:mfe_6sites}(a--d), demonstrate the effectiveness of the standard full MFE, where phonon dynamics is governed by Eq.~\eqref{P-evol} and Eq.~\eqref{Q-evol}. Compared to the standard CE method, MFE performs well overall aside from the failure to produce strictly positive spectral functions. MFE describes some aspects of the spectra better than CE does at $k=\pi$ at both $T=0.2$ (Fig.~\ref{fig:mfe_6sites}c) and $T=1.0$ (Fig.~\ref{fig:mfe_6sites}d) by reproducing the split peaks that CE misses. These peaks arise from oscillations associated with the finite spectrum of a small system, and tend to disappear for large $N$. At $k=0$ and low temperature (Fig.~\ref{fig:mfe_6sites}a), MFE reasonably captures the main peak, although the CE is the most reliable approach, resolving the structured spectral features more accurately. Meanwhile, at $k=0$ and high temperature (Fig.~\ref{fig:mfe_6sites}b), both MFE and CE are able to describe the general structure of the spectral function, consistent with the VD results.

When MFE is simplified by removing the full dynamical treatment of phonons, the resulting spectral features are notably altered. The middle panels, Fig. \ref{fig:mfe_6sites}(e--h), illustrate the case of frozen MFE, where the bath variables are fixed at their initial values throughout the propagation. Unlike standard MFE, frozen MFE produces a broad peak across all regimes of $k$-points ($k=0$ and $k=\pi$) and temperatures ($T=0.2$ and $T=1.0$). While it somewhat accurately captures the positions of the quasiparticle peaks, it fails to replicate the detailed spectral features. 

The lower panels, Fig.~\ref{fig:mfe_6sites}(i--l), show the results of MFE without the back-reaction force. In this approach, the bath variables are propagated as usual, but the average force from the electronic wavefunction in Eq.~\eqref{P-evol} and Eq.~\eqref{Q-evol} is removed. This simplification introduces inaccuracies in peak locations. The sensitivity of spectral properties to the back-reaction force lies in contrast to the insensitivity of transport coefficients in the same model.~\cite{WangJCP2011} These observations underscore the importance of retaining the full dynamical treatment of phonons in terms of the accuracy of methods based on MFE.

\begin{figure}
    \centering
    \includegraphics[width=\columnwidth]{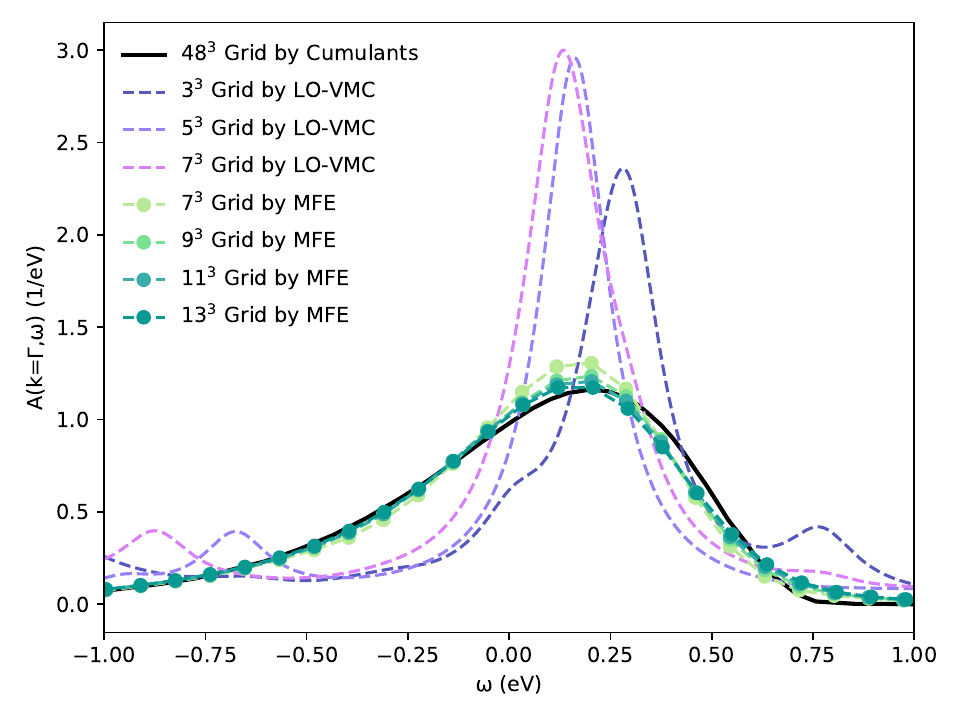}
    \caption{(Color online) Spectral function at the $\Gamma$-point for an \textit{ab initio} model of LiF computed using MFE, the CE, and variational Monte Carlo with only longitudinal optical phonons (LO-VMC) for varying $k$-grid sizes at zero temperature. CE data were digitally extracted from Nery \textit{et al.}~\cite{Nery2018} The broadening parameter is $0.01$ eV. Further details of the model can be found in Ref.~\onlinecite{robinson2025ab}.}
    \label{fig:lif_at_gamma}
\end{figure}

With the reasonable performance of the full standard MFE established, we apply the method to the \textit{ab initio} model of LiF (Eq.~\eqref{eq:H_lif}) at zero temperature.
As shown in Fig.~\ref{fig:lif_at_gamma}, MFE reproduces the CE result at the $\Gamma$-point as reported in Ref.~\onlinecite{Nery2018} as the grid size is increased. This result suggests that MFE essentially achieves convergence in this example for a relatively small grid size ($7^3$), producing results comparable to the CE method obtained at a much larger grid ($48^3$).

In contrast, the near-exact variational Monte Carlo (VMC) \cite{mahajan2024structure} (see SM for details of the method) is constrained by the expense of large grid sizes. The VMC spectral function for grid sizes of $3^3$, $5^3$, and $7^3$ $k$-points is shown, revealing significant deviations from both MFE and CE results. Finite-size effects appear to be larger within the VMC calculations than MFE calculations. However, fully converged results will still likely produce a sharper VMC spectral function. We therefore conclude that the discrepancies shown by VMC results reflect true differences between the approximate and near-exact results, and are not associated with differences in grid sizes. Interestingly, the VMC results appear to have features that are absent in the cumulant results but appear in the Dyson-Migdal formalism.~\cite{Nery2018} Both VMC and Dyson-Migdal spectra show a sharp peak centered at $\omega\approx0.2$ eV and a broad shoulder at the lower frequencies ($\omega\approx-1$ to $0$ eV).

\section{Summary and Conclusions}\label{sec:summary}

In this work, we have compared distinct semiclassical approaches, namely the standard mean field Ehrenfest theory (MFE) and the more recently formulated MASH approach, by considering the 1-particle Green's function for electron-phonon systems such as the one-dimensional Holstein model. 
We compare these approaches to exact results as determined by variational exact diagonalization and variational quantum Monte Carlo. Our comparisons are aided by analytical calculations in the simplified limit of zero hopping.

Focusing on the one-dimensional Holstein model, we find that both MFE and MASH produce similar spectra for $k$ values at the band minimum and maximum when the number of sites is large. These results are largely consistent with exact diagonalization results.
In line with this, Runeson \textit{et al.}~\cite{Runeson2024b} report semi-quantitative, instead of quantitative, agreements between MFE and MASH in computing charge transport in the Su-Schrieffer-Heeger model.
However, for small periodic lattices, we find that MFE is capable of reproducing sharp features in the spectral function more accurately than MASH, albeit at the expense of some small, unphysical spectral regions. This negative spectral weight appears to be suppressed in our formulation of MASH for standard model parameters.

Within MFE alone, we examined the role played by the treatment of the phonon dynamics, finding that unlike the case for transport problems in some regimes~\cite{WangJCP2011}, inclusion of the back-reaction force notably increases the accuracy of the method. We have also applied MFE to an \textit{ab initio} model of the spectral function for a valence hole in LiF, and compare the results to near-exact real-frequency variational quantum Monte Carlo calculations. We find that MFE quantitatively agrees with previous cumulant calculations, and that both of these approximate methods produce spectra which appear to be notably broader than that found by Monte Carlo calculations performed on somewhat smaller grids.

Overall, our work illustrates that semiclassical methods offer an attractive and inexpensive means to estimate spectral properties of electron-phonon coupled systems. These methods are competitive with, and sometimes more accurate than, other approximate methods which have been used previously to calculate spectral functions in polaronic systems, and are easy to adapt for calculations in complex \textit{ab initio} models of solids.

{\noindent}{\bf Supplementary Material}:
See the supplementary materials for 1) correlation functions in the mixed quantum-classical limit; 2) estimator of electronic wavefunction coefficients in the Multi-State Mapping Approach to Surface Hopping method; 3) expression for the correlation function in the pure-state decomposition method; 4) analytical expressions for the spectral functions of the Holstein model at $J=0$ and zero temperature; and 5) details of the Variational Monte Carlo with neural quantum states method used for computing the LiF spectral function.

{\noindent}{\bf Code availability}: Code and results in this work can be found at \url{https://github.com/nguye66h/mqc-specfunc}.

{\noindent}{\bf Acknowledgements}:
We thank David Manolopoulos for a discussion and for sending us Ref.~\onlinecite{manounpub}.
The work of HN and DRR was performed with support from the U.S.\ Department of Energy, Office of Science, Office of Advanced Scientific Computing Research, Scientific Discovery through Advanced Computing (SciDAC) program, under Award No. DE-SC0022088.  This work used resources of the National Energy Research Scientific Computing Center (NERSC), a U.S. Department of Energy Office of Science User Facility located at Lawrence Berkeley National Laboratory, operated under Contract No. DE-AC02-05CH11231.  Some of this work used TAMU FASTER at the Texas A\&M University through allocation  PHY230021 from the Advanced Cyberinfrastructure Coordination Ecosystem: Services \& Support (ACCESS) program, which is supported by National Science Foundation grants \#2138259, \#2138286, \#2138307, \#2137603, and \#2138296.  

{\noindent}{\bf Author Declarations}:

{\noindent}{\bf Conflict of Interest}:
The authors have no conflicts to disclose.

{\noindent}{\bf Author Contributions}:
\textbf{Haimi Nguyen}: Conceptualization (equal); Data curation (lead); Investigation (lead); Methodology (equal); Software (equal); Visualization (lead); Writing - original draft preparation (lead); Writing - reviewing \& editing (lead).
\textbf{Arkajit Mandal}: Conceptualization (equal); Investigation (supporting); Methodology (equal); Software (equal); Visualization (supporting); Project administration (supporting); Resources (supporting); Supervision (supporting); Writing - original draft preparation (supporting); Writing - reviewing \& editing (supporting).
\textbf{Ankit Mahajan}: Conceptualization (equal); Data curation (supporting); Investigation (supporting); Methodology (supporting); Software (supporting); Validation (lead); Writing - original draft preparation (supporting); Writing - reviewing \& editing (supporting).
\textbf{David Reichman}: Conceptualization (equal); Funding acquisition (lead); Project administration (lead); Resources (lead); Supervision (lead); Writing - original draft preparation (supporting); Writing - reviewing \& editing (supporting).

{\noindent}{\bf Data Availability}:
The data that support the findings of this study are available upon reasonable request.

\bibliographystyle{aipnum4-2}
\bibliography{ref}

\end{document}


\begin{center}
\Large Supplementary Materials for ``Mixed Quantum-Classical Methods for Polaron Spectral Functions''
\end{center}

\vspace*{2em}

This supplement is divided into 5 parts:
\begin{itemize}
    \item Correlation functions in the mixed quantum-classical limit
    \item Estimator of electronic wavefunction coefficients in the Multi-state Mapping Approach to Surface Hopping (MASH) method
    \item Expression for the correlation function in the pure-state decomposition method
    \item Analytical expressions for the spectral functions of the Holstein model at $J=0$ and zero temperature
    \item Variational Monte Carlo with neural quantum states
\end{itemize}

\section{General Expressions for the Correlation Functions in the Mixed Quantum-Classical Methods in the Single Excitation Basis}

In this section, we derive an expression for the correlation function in general mixed quantum-classical methods, where the phonon momenta and positions are classical variables, and electronic states are treated quantum mechanically. 
We start from the Hamiltonian of the one-dimensional Holstein model in the single excitation basis,
\begin{align}
    H&=H_S+H_{SB}+H_{B}\\
     &=\sum_{n=0}^{N-1} -J\left(\ketbra{n+1}{n} + \ketbra{n}{n+1}\right)+ g\ketbra{n}{n} (b^\dagger_n + b_n)+ \omega_0 b^\dagger_n b_n.\label{eqn_SM:realspace_Holstein_Hamiltonian}
\end{align}
The correlation function has the form of
\begin{align}
    C_k(t) &= \frac{1}{Z}\operatorname{Tr}\left(e^{-\beta H} e^{i H t} \hat{c}_k e^{-i H t}  \hat{c}_k^\dagger \right)\\
                &=\frac{1}{Z} \sum_n \bra{n}_{\text{ph}} \otimes \bra{\emptyset}  e^{-\beta H}  e^{i H t} \hat{c}_k e^{-i H t}  \hat{c}_k^\dagger \ket{\emptyset} \otimes \ket{n}_{\text{ph}} \\
                &=\frac{1}{Z_B} \sum_n \bra{n}_{\text{ph}} e^{-\beta H_{B}} e^{i H_B t}\otimes \bra{\emptyset}   \hat{c}_k e^{-i H t}  \hat{c}_k^\dagger \ket{\emptyset} \otimes \ket{n}_{\text{ph}}\\
                &=\frac{1}{Z_B} \sum_n \bra{n}_{\text{ph}} e^{-\beta H_{B}} e^{i H_B t}\otimes \bra{k}    e^{-i H t}  \ket{k} \otimes \ket{n}_{\text{ph}}, \label{eqn_SM:exact_correlation_function}
\end{align}
where $\ket{\emptyset}=\otimes_{i=0}^{N-1}\ket{\emptyset}_i$ is the vacuum state for $N$ sites, while $\ket{n}_\text{ph}=\otimes_{i=0}^{N-1}\ket{n}_{\text{ph},i}$ is the occupation number state for $N$ local baths associated with $N$ sites.
The above derivation is based on the fact that
\begin{align}
    e^{a H } \ket{\emptyset} \otimes \ket{n}_{\text{ph}} &= (1 + a (H_S+H_{SB}+H_B) + \frac{1}{2}\left[a(H_S+H_{SB}+H_B) \right]^2+\ldots) \ket{\emptyset} \otimes \ket{n}_{\text{ph}} \\
    &=(1 + a H_B + \frac{1}{2} \left( a H_B  \right)^2+\ldots) \ket{\emptyset} \otimes \ket{n}_{\text{ph}} \\
    &= \ket{\emptyset} \otimes e^{a H_B } \ket{n}_{\text{ph}},
\end{align}
since $H_S$, $H_{SB}$, and any products of these operators acting on the vacuum state of the system give zero.
In a mixed quantum-classical method, since $H_B$ becomes a function of the bath momenta and positions, $P$ and $Q$, respectively, Eq.~\eqref{eqn_SM:exact_correlation_function} is now
\begin{align}
    C_k(t) &=\frac{1}{Z_B} \int d\vec{P}_0 d\vec{Q}_0 e^{-\beta E_{B}(\vec{P}_0, \vec{Q}_0)} \prod_{i=0}^{M-1} e^{i E_{B}(\vec{P}_i, \vec{Q}_i) \Delta t} \bra{k}   \prod_{i=0}^{M-1} e^{-i (H_S + H_{SB}) \Delta t} e^{-i E_{B}(\vec{P}_i, \vec{Q}_i) \Delta t}  \ket{k}\\
    &=\frac{1}{Z_B} \int d\vec{P}_0 d\vec{Q}_0 e^{-\beta E_{B}(\vec{P}_0, \vec{Q}_0)} \bra{k}   \prod_{i=0}^{M-1} e^{-i (H_S + H_{SB}) \Delta t}  \ket{k}\\
    &=\frac{1}{Z_B} \int d\vec{P}_0 d\vec{Q}_0 e^{-\beta E_{B}(\vec{P}_0, \vec{Q}_0)} \bra{k}   e^{-i (H_S + H_{SB}) t}  \ket{k}\\
    &= \left \langle \bra{k}   e^{-i (H_S + H_{SB}) t}  \ket{k} \right \rangle_\text{MQC},
\end{align}
where $M=\frac{t}{\Delta t}$. Note that $H_{SB}$ is dependent on $\vec{P}(t)$ and $\vec{Q}(t)$ as parameters.

In mean field Ehrenfest (MFE), the correlation function is the amplitude $c_k(t)$ of state $\ket{k}$, calculated when the initial wavefunction is $\ket{k}$ and averaged over the trajectories of the classical variables. In contrast, in MASH, the correlation function is represented by the electronic coefficient estimator $W_{kk}(t)$ averaged over the trajectories of the classical phonon variables and initial wavefunction coefficients.

\section{Estimator of electronic wavefunction coefficients in MASH}

In this section, we derive an expression to estimate the coefficients of the electronic wavefunctions in MASH. To do so, we establish a connection between elements of the propagator in the original system, $\expecth{k}{e^{-iHt}}{i}$, and elements of the density matrix of the expanded system by introducing a fictitious state $\ket{\phi}$, which is uncoupled from the original system by the Hamiltonian. Note that the state $\ket{i}$ here corresponds to the state $\ket{\psi_0}$ in Eq.~(25) of the main text. Letting the initial wavefunction be
\begin{align}
    \ket{\tilde{i}}=m_\phi \ket{\phi} + m_i \ket{i},
\end{align}
where $m_\phi$ and $m_i$ are real numbers for simplicity, and $m_\phi^2+m_i^2=1$, we consider an element of the density matrix,
\begin{align}
    \rho_{k\phi} (t)&= \expecth{k}{e^{-iHt}}{\Psi(0)}  \expecth{\Psi(0)}{e^{iHt}}{\phi}\\
    &=\langle k | e^{-iHt} (m_\phi \ket{\phi} + m_i \ket{i})  (\bra{\phi} m_\phi  + \bra{i} m_i)e^{iHt}\ket{\phi} \label{eqn:rho_1} \\
    &=  m_i  m_\phi \expecth{k}{e^{-iHt}}{i}. \label{eqn:rho_2}
\end{align}
The derivation of Eq.~\eqref{eqn:rho_2} from Eq.~\eqref{eqn:rho_1} relies on the fact that the Hamiltonian does not act on the $\ket{\phi}$ state in any way.
From Ref.~\onlinecite{Runeson2023},
\[\langle O\rangle=\sum_{nm}O_{nm}\rho_{mn}\rightarrow\sum_{nm}O_{nm}\left[\alpha_Nc^*_nc_m+\beta_N\delta_{nm}\right],\]
estimators for elements in electronic observables are complex conjugates to estimators for elements in density matrices. $\alpha_N$ and $\beta_N$ are constants that we later define at Eq.~\eqref{eqn:alpha} and Eq.~\eqref{eqn:beta}, and $c_n$ is the amplitude coefficient for state $\ket{n}$.
From Eq.~\eqref{eqn:rho_2}, it is now clear that
\begin{align}
    \left \langle \expecth{k}{e^{-iHt}}{i} \right \rangle_\text{MQC}=\frac{\left \langle \rho_{k\phi}(t) \right \rangle_\text{MQC}}{m_i  m_\phi}=\frac{\left \langle\alpha_{N} c_k (t) c^*_\phi (t) \right \rangle_\text{MASH}}{m_i  m_\phi}=\left \langle W_{ki} (t) \right \rangle_\text{MASH}. \label{eqn:estimator}
\end{align}
where 
\begin{align}
    \langle O (t) \rangle_\text{MQC}=\int d\vec{P}_0 d\vec{Q}_0 \rho_W(\beta,\vec{P}_0,\vec{Q}_0) O(\vec{P}_0,\vec{Q}_0,t)
\end{align}
and
\begin{align}
    \langle O (t) \rangle_\text{MASH}=\int d\vec{P}_0 d\vec{Q}_0 \int d\vec{c}_0 \rho_W(\beta,\vec{P}_0,\vec{Q}_0)\rho_M(\vec{c}_0) O(\vec{P}_0,\vec{Q}_0,\vec{c}_0,t).
\end{align}
$\rho_W$ and $\rho_M$ are the Wigner and MASH initial density functions for the bath variables and electronic coefficients respectively. $W_{ki}(t)$ is the estimator for $\expecth{k}{e^{-iHt}}{i}$ that we need to determine. The $N$ in $\alpha_{N}$ \textit{includes} the auxiliary state.

\begin{figure}
    \centering
    \includegraphics[width=1.0\linewidth]{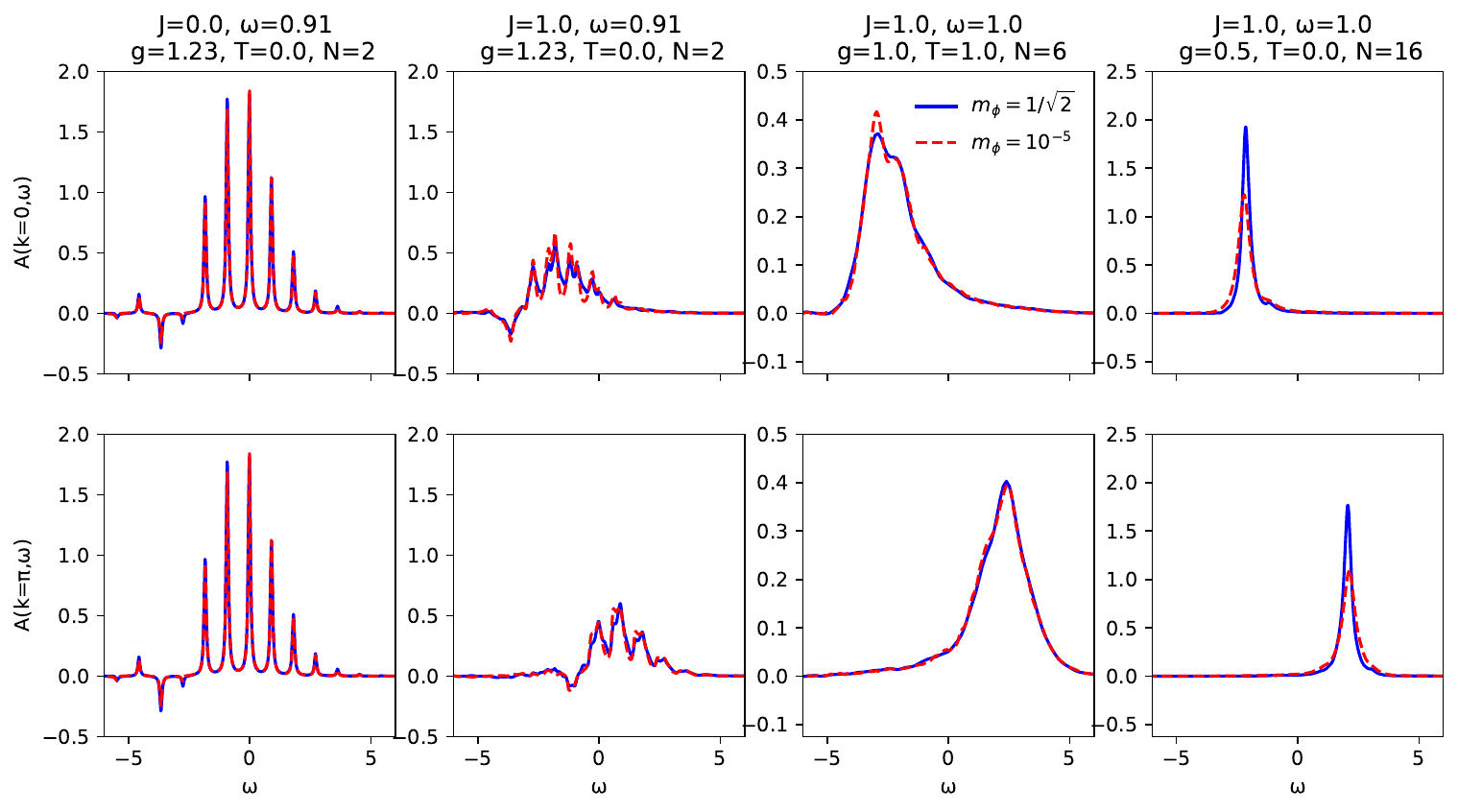}
    \caption{Spectral functions computed by MASH in different regimes for $m_\phi=\frac{1}{\sqrt{2}}$ (blue curve) and $m_\phi=10^{-5}$ (red dashed curve).}
    \label{fig:compare_mphi}
\end{figure}

In order to sample from $\rho_M(\vec{c}_0)$, one needs to establish an orthonormal basis, one of whose basis vectors is the state we want to start from. For our case, we want to start with state $\ket{\tilde{i}}$. With the original orthonormal basis set \{$\ket{i}$\}, our new orthonormal basis is now
\begin{align}
    \ket{\tilde{i}}&=m_i \ket{i} + m_\phi \ket{\phi}, \label{eqn:tilde_i}\\
    \ket{\tilde{\phi}}&=m_\phi \ket{i} - m_i \ket{\phi},\\
    \ket{\tilde{j}}&=\ket{j},
\end{align}
where states $|j\rangle$ denote the basis vectors orthogonal to $|i\rangle$, and together they span the original Hilbert space. 
We take $m_i$ and $m_\phi$ as real numbers. In the limit of $m_i \rightarrow 1$ and $m_\phi \rightarrow 0$, we recover the original basis.
Here, we take $\rho_M(\vec{c}_0)$ to be the focused initial sampling~\cite{Runeson2024}, where the magnitudes of the wavefunction coefficients are fixed to predetermined values, and only the phases are sampled. However, this derivation also applies to other sampling schemes.
The initial wavefunction focused on state $\ket{\tilde{i}}$ (Eq.~\eqref{eqn:tilde_i}) is
\begin{align}
    \ket{\psi(0)}=\sqrt{\frac{1+\beta_N}{\alpha_N}}e^{i\theta_{\tilde{i}}} \ket{\tilde{i}}+\sqrt{\frac{\beta_N}{\alpha_N}} e^{i\theta_{\tilde{\phi}}} \ket{\tilde{\phi}}
    +\sum_{j\neq i} \sqrt{\frac{\beta_N}{\alpha_N}} e^{i\theta_j} \ket{j},
\end{align}
where
\begin{align}
    \alpha_N&=\frac{N-1}{\sum_{k=1}^N \frac{1}{k}-1}, \label{eqn:alpha} \\
    \beta_N&=\frac{\alpha_N-1}{N}. \label{eqn:beta}
\end{align}
$N$ represents the total system size, \textit{including} the fictitious state. The wavefunction can be rewritten in the expanded basis as
\begin{align}
|\psi(0)\rangle&=\left(m_\phi\sqrt{\frac{1+\beta_N}{ \alpha_N}} e^{i \theta_{\tilde{i}}}-m_i\sqrt{\frac{\beta_N}{ \alpha_N}} e^{i \theta_{\tilde{\phi}}}\right)|\phi\rangle +\left(m_i\sqrt{\frac{1+\beta_N}{ \alpha_N}} e^{i \theta_{\tilde{i}}}+m_\phi\sqrt{\frac{\beta_N}{ \alpha_N}} e^{i \theta_{\tilde{\phi}}}\right)\ket{i} +\sum_j \sqrt{\frac{\beta_N}{\alpha_N}} e^{i \theta_{j}}|j\rangle.
\end{align}
Since the state $\ket{\phi}$ is completely uncoupled from the original system, propagating $\ket{\psi}$ gives
\begin{align}
|\psi(t)\rangle &=\left(m_\phi\sqrt{\frac{1+\beta_N}{ \alpha_N}} e^{i \theta_{\tilde{i}}}-m_i\sqrt{\frac{\beta_N}{ \alpha_N}} e^{i \theta_{\tilde{\phi}}}\right)|\phi\rangle +\left(m_i\sqrt{\frac{1+\beta_N}{ \alpha_N}} e^{i \theta_{\tilde{i}}}+m_\phi\sqrt{\frac{\beta_N}{ \alpha_N}} e^{i \theta_{\tilde{\phi}}}\right)\sum_{k'}\gamma_{k'i}(t)\ket{k'} +\sum_{j} \sqrt{\frac{\beta_N}{\alpha_N}} e^{i \theta_j}\sum_{k'}\gamma_{k'j}(t)|k'\rangle,
\end{align}
where
\begin{align}
    \gamma_{mn}(t)=\langle m | e^{-iHt} | n \rangle.
\end{align}
We can now expand Eq.~\eqref{eqn:estimator}:
\begin{align}
    \frac{\left \langle\alpha_N c_k (t) c^*_\phi (t) \right \rangle_\text{MASH}}{m_i  m_\phi}
    &=\frac{\alpha_N \left \langle \left( \left(m_i\sqrt{\frac{1+\beta_N}{ \alpha_N}} e^{i \theta_{\tilde{i}}}+m_\phi\sqrt{\frac{\beta_N}{ \alpha_N}} e^{i \theta_{\tilde{\phi}}}\right)\gamma_{ki}(t) +  \sum_{j} \sqrt{\frac{\beta_N}{\alpha_N}} e^{i \theta_j} \gamma_{kj}(t) \right) \left(m_\phi\sqrt{\frac{1+\beta_N}{ \alpha_N}} e^{-i \theta_{\tilde{i}}}-m_i\sqrt{\frac{\beta_N}{ \alpha_N}} e^{-i \theta_{\tilde{\phi}}}\right) \right \rangle_\text{MASH}}{m_i  m_\phi}\\
    &=\alpha_N \left \langle \gamma_{ki}(t) \left(\frac{1+\beta_N}{ \alpha_N} -\frac{\beta_N}{ \alpha_N}+\sqrt{\frac{1+\beta_N}{ \alpha_N}}\sqrt{\frac{\beta_N}{ \alpha_N}}\left( \frac{m_\phi}{m_i}e^{-i \theta_{\tilde{i}}}e^{i \theta_{\tilde{\phi}}}-\frac{m_i}{m_\phi}e^{i \theta_{\tilde{i}}}e^{-i \theta_{\tilde{\phi}}} \right)\right) \right \rangle_\text{MASH} \\
    &  + \alpha_N\left \langle  \sum_{j} \gamma_{kj}(t) \sqrt{\frac{\beta_N}{\alpha_N}} e^{i \theta_{j}} \left(\frac{1}{m_i}\sqrt{\frac{1+\beta_N}{ \alpha_N}} e^{-i \theta_{\tilde{i}}}-\frac{1}{m_\phi}\sqrt{\frac{\beta_N}{ \alpha_N}} e^{-i \theta_{\tilde{\phi}}}\right) \right \rangle_\text{MASH} \nonumber \\
    &=\left \langle   \gamma_{ki}(t) \right \rangle_\text{MASH}.
\end{align}
The previous derivation works because any term that averages over random phases will equate zero. That leaves us with a simple final expression.
On the other hand,
\begin{align}
    \left \langle\frac{ c_k(t) }{ c_i(0) }\right \rangle_\text{MASH}&=\left \langle\frac{ \left(m_i\sqrt{\frac{1+\beta_N}{ \alpha_N}} e^{i \theta_{\tilde{i}}}+m_\phi\sqrt{\frac{\beta_N}{ \alpha_N}} e^{i \theta_{\tilde{\phi}}}\right)\gamma_{ki}(t)+\sum_{j} \sqrt{\frac{\beta_N}{\alpha_N}} e^{i \theta_{j}}\gamma_{kj}(t) }{\left(m_i\sqrt{\frac{1+\beta_N}{ \alpha_N}} e^{i \theta_{\tilde{i}}}+m_\phi\sqrt{\frac{\beta_N}{ \alpha_N}} e^{i \theta_{\tilde{\phi}}}\right) }\right \rangle_\text{MASH}\\
    &=\left \langle  \gamma_{ki}(t) \right \rangle_\text{MASH}.
\end{align}
We can now establish the MASH estimator for $\expecth{k}{e^{-iHt}}{i}$ as
\begin{align}
    W_{ki}(t)= \frac{c_k(t)}{c_i(0)}.
\end{align}
In our MASH algorithm, even when $\ket{\phi}$ plays no part in the MASH dynamics, it is clear that the values of $m_\phi$ and $m_k$ affect the relative populations between states $|n\rangle$, leading to potentially different selections of the adiabatic surfaces acting on the MASH force throughout the propagation. One clear example for this is Eq.~\eqref{eq:MASHwf_inrspace}, which we later show for the dynamics in the exact case of $J=0$. However, from our empirical observations, this effect is negligible on the resulting spectral functions. Fig.~\ref{fig:compare_mphi} shows a few representative cases from the main paper to demonstrate the robustness of the spectral functions against the change in the value of $m_\phi$.

\section{Expression for the correlation function in the pure-state decomposition method}

\begin{figure}
    \centering
    \includegraphics[width=1.0\linewidth]{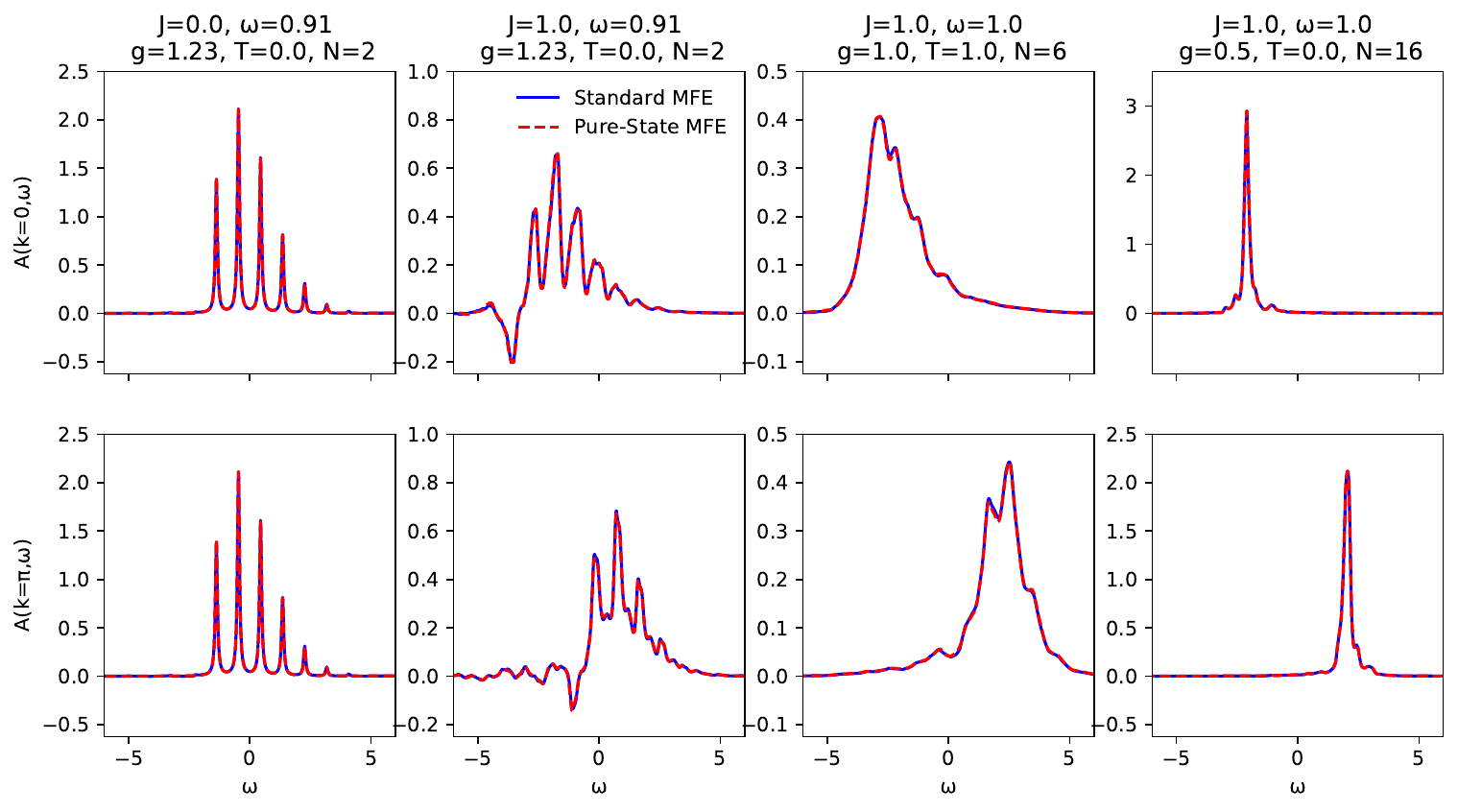}
    \caption{Spectral functions computed by MFE in different regimes using the standard approach (blue curve) and the pure-state decomposition approach (red dashed curve).}
    \label{fig:compare_MFE}
\end{figure}

\begin{figure}
    \centering
    \includegraphics[width=1.0\linewidth]{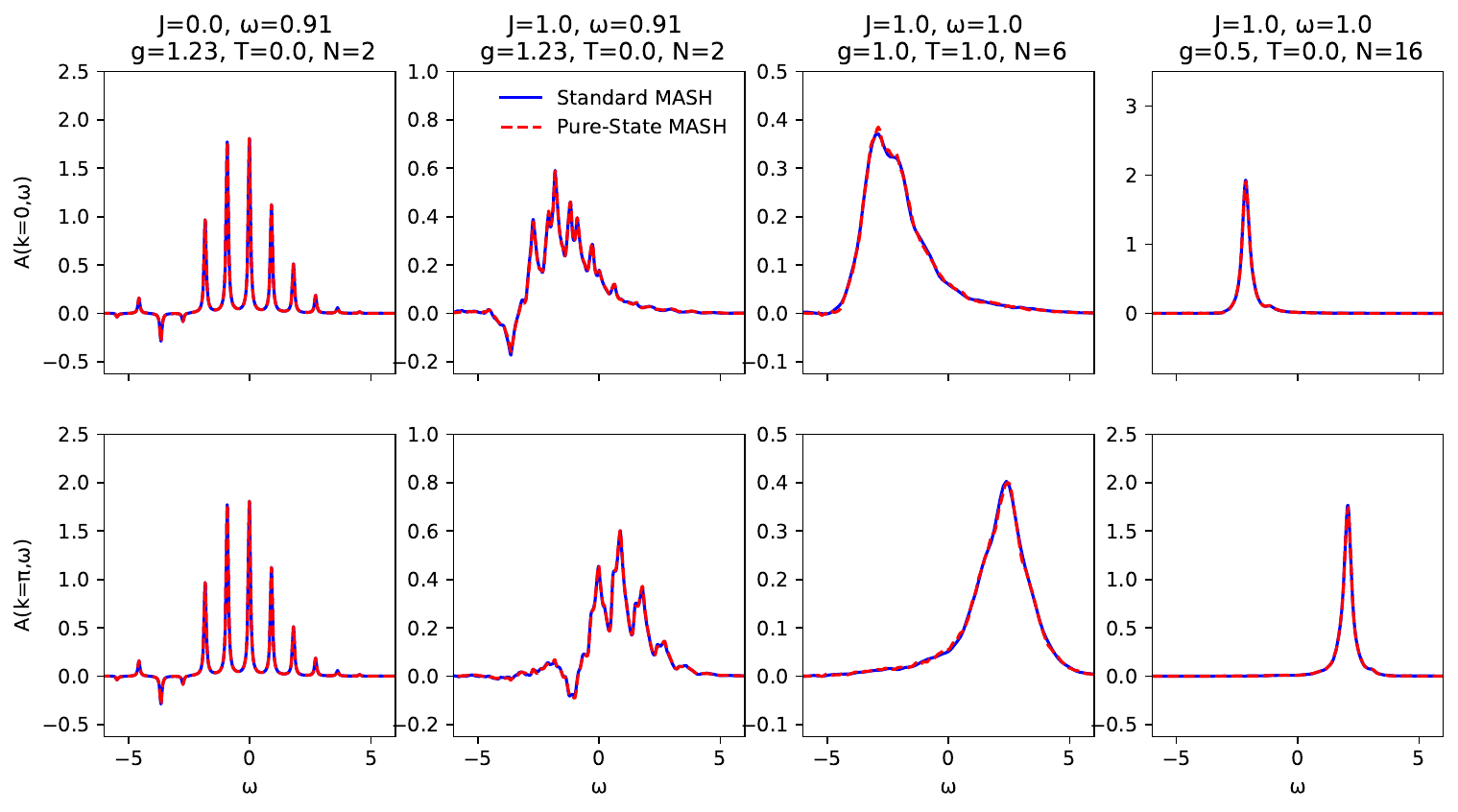}
    \caption{Spectral functions computed by MASH in different regimes using the standard approach (blue curve) and the pure-state decomposition approach (red dashed curve).}
    \label{fig:compare_MASH}
\end{figure}

In the single-particle projection basis, we can rewrite $c_k\rightarrow|\emptyset\rangle\langle k|$ to get the following expressions for the correlation function,
\begin{align}
    C_k(t)&=\langle c_k (t) c^\dagger_k(0) \rangle\\
    &=\left \langle e^{iHt}|\emptyset\rangle \langle k |e^{-iHt} | k\rangle\langle\emptyset| \right \rangle\\
    &=\frac{\Tr_B\left[ e^{-\beta H_B}\langle \emptyset| \sum_{j=0}^3 w_j e^{iHt}|j\rangle\langle j|e^{-iHt}\sum_{j'=0}^3 w_{j'} |\tilde{j'}\rangle\langle \tilde{j'}|\emptyset \rangle \right]}{\Tr_B[e^{-\beta H_B}]}\\
    &=\frac{\Tr_B\left[ e^{-\beta H_B}\langle \emptyset| \sum_{j=0}^3 w_j |j(-t)\rangle\langle j(-t)|\sum_{j'=0}^3 w_{j'} |\tilde{j'}\rangle\langle \tilde{j'}|\emptyset \rangle \right]}{\Tr_B[e^{-\beta H_B}]}\\
    &=\frac{1}{\Tr_B\left[e^{-\beta H_B}\right]}\Tr_B\left[e^{-\beta H_B}\frac{1}{\sqrt{2}}(\langle0|+\langle2|)\sum_{j=0}^3 w_j |j(-t)\rangle\langle j(-t)|\sum_{j'=0}^3 w_{j'} |\tilde{j'}\rangle\langle \tilde{j'}|\frac{1}{\sqrt{2}}(|\tilde{0}\rangle-|\tilde{2}\rangle)\right].\label{eqn:mash_pure_state}
\end{align}
From the definition of the pure-state decomposition,
\begin{align}
    \ketbra{a}{b}&=\sum_{j=0}^3w_j\ketbra{j}{j},\\
    \ket{j}&=\frac{1}{\sqrt{2}}\left(\frac{\ket{a}}{\sqrt{\braket{a}{a}}}+e^{ij\frac{\pi}{2}}\frac{\ket{b}}{\sqrt{\braket{b}{b}}}\right),\\
    w_j&=\frac{\sqrt{\braket{a}{a}\braket{b}{b}}}{2}e^{ij\frac{\pi}{2}},
\end{align}
we know that
\begin{alignat}{4}
    w_0 = \frac{1}{2},     &\quad 
    w_1 = \frac{i}{2},      &\quad 
    w_2 = \frac{-1}{2},    &\quad 
    w_3 = \frac{-i}{2},
\end{alignat}
\begin{alignat}{4}
    \ket{0}=\frac{1}{\sqrt{2}}(\ket{\emptyset}+\ket{k}), &\quad
    \ket{1}=\frac{1}{\sqrt{2}}(\ket{\emptyset}+i\ket{k}),&\quad
    \ket{2}=\frac{1}{\sqrt{2}}(\ket{\emptyset}-\ket{k}),&\quad
    \ket{3}=\frac{1}{\sqrt{2}}(\ket{\emptyset}-i\ket{k}),
\end{alignat}
\begin{alignat}{4}
    \ket{\tilde{0}}=\frac{1}{\sqrt{2}}(\ket{k}+\ket{\emptyset}),&\quad
    \ket{\tilde{1}}=\frac{1}{\sqrt{2}}(\ket{k}+i\ket{\emptyset}),&\quad
    \ket{\tilde{2}}=\frac{1}{\sqrt{2}}(\ket{k}-\ket{\emptyset}),&\quad
    \ket{\tilde{3}}=\frac{1}{\sqrt{2}}(\ket{k}-i\ket{\emptyset}).
\end{alignat}
Now, with the relations
\begin{alignat}{4}
    \ket{\tilde{0}}=\ket{0},&\quad
    \ket{\tilde{1}}=i\ket{3},&\quad
    \ket{\tilde{2}}=-\ket{2},&\quad
    \ket{\tilde{3}}=-i\ket{1},&\quad
    w_1\ket{3}+w_3\ket{1}=w_0(\ket{0}-\ket{2}),
\end{alignat}
Eq.~\eqref{eqn:mash_pure_state} becomes
\begin{align}
    C_k(t)
    &=\frac{\Tr_B \left[ e^{-\beta H_B} \sum_{j=0}^3 w_jw_0 \left(\braket{0}{j(-t)}\braket{j(-t)}{0}+\braket{2}{j(-t)}\braket{j(-t)}{0}\right) \right]}{\Tr_B [ e^{-\beta H_B}]}\nonumber\\
    &+\frac{\Tr_B \left[ e^{-\beta H_B} \sum_{j=0}^3 w_jw_2\left(\braket{0}{j(-t)}\braket{j(-t)}{2}+\braket{2}{j(-t)}\braket{j(-t)}{2} \right) \right]}{\Tr_B [ e^{-\beta H_B}]} \nonumber\\
    &=\frac{\Tr_B \left[ e^{-\beta H_B} \sum_{j=0}^3 w_jw_0\left(\braket{0(t)}{j}\braket{j}{0(t)} +\braket{2(t)}{j}\braket{j}{0(t)}\right)\right]}{\Tr_B [ e^{-\beta H_B}]}\nonumber\\
    &+\frac{\Tr_B \left[ e^{-\beta H_B} \sum_{j=0}^3 w_jw_2\left(\braket{0(t)}{j}\braket{j}{2(t)} +\braket{2(t)}{j}\braket{j}{2(t)}\right)\right]}{\Tr_B [ e^{-\beta H_B}]}\\
    &=\frac{\Tr_B \left[ e^{-\beta H_B} \sum_{j=0}^3 w_jw_0\left(\braket{j}{0(t)}\braket{0(t)}{j} -\braket{j}{2(t)}\braket{2(t)}{j}+i\braket{j}{+i(t)}\braket{+i(t)}{j}-i\braket{j}{-i(t)}\braket{-i(t)}{j}\right)\right]}{\Tr_B [ e^{-\beta H_B}]}, \label{eqn:pure-state-corr}
\end{align}
where
\begin{align}
    \ket{+i}=\frac{1}{\sqrt{2}}(\ket{0}+i\ket{2}),\\
    \ket{-i}=\frac{1}{\sqrt{2}}(\ket{0}-i\ket{2}).
\end{align}
Eq.~\eqref{eqn:pure-state-corr} allows one to compute MASH spectral function from the standard population estimator established in Ref.~\onlinecite{Runeson2023}. We compare numerical evaluations of the MFE (Fig.~\ref{fig:compare_MFE}) and MASH (Fig.~\ref{fig:compare_MASH}) spectral functions by Eq.~(17) (standard approach) and Eq.~(45) (pure-state decomposition) of the main text. We conclude that our method and the pure-state decomposition method give indistinguishable numerical results.

\section{Analytical Expressions for Spectral Functions of the Holstein Model at $J=0$ and Zero Temperature}
\subsection{The Exact Solution}

Given the Hamiltonian in Eq.~\eqref{eqn_SM:realspace_Holstein_Hamiltonian} with $J=0$, 
\begin{align}
    H = \sum_n \, g \omega_0 \ketbra{n}{n} (b_n + b_n^\dagger) + \omega_0 b_n^\dagger b_n,
\end{align}
we define the unitary operator,
\begin{align}
    U = \exp \left[\sum_n g \ketbra{n}{n} (b^\dagger_n - b_n)\right].
\end{align}
At zero temperature,
\begin{align}
    C_k(t) 
                &= \frac{1}{Z_B} \bra{\emptyset}_{\text{ph}} \otimes  \bra{k} e^{-i H t}  \ket{k} \otimes \ket{\emptyset}_{\text{ph}} \label{exactcorr_2}\\ 
    &=\frac{1}{Z_B} \bra{\emptyset}_{\text{ph}} \otimes \sum_n \bra{n} \frac{e^{-ikn}}{\sqrt{N}} U^\dagger U e^{-iHt} U^\dagger U \sum_{n'}\frac{e^{ikn'}}{\sqrt{N}}\ket{n'} \otimes \ket{\emptyset}_{\text{ph}},
\end{align}
where $\ket{\emptyset}_\text{ph}=\otimes_{i=1}^N\ket{\emptyset}_{\text{ph},i}$ is the vacuum state for $N$ local baths associated with $N$ sites, and $\ket{n}=\ket{\emptyset}_0 \otimes \cdots \otimes \ket{1}_n \otimes \cdots \otimes \ket{\emptyset}_{N-1}$ means an occupation of one electron on site $n$ while the other sites remain empty.
We know that
\begin{align}
    U \ket{n} \otimes \ket{\emptyset}_{\text{ph}} &= \ket{n} e^{g(b^{\dagger}_n-b_n)} \ket{\emptyset}_{\text{ph}} \\
    &=\ket{n} \otimes \ket{g}_{n,\text{ph}} \otimes_{m \neq n} \ket{\emptyset}_{m,\text{ph}}\\
    &=\ket{n} \otimes e^{-g^2/2} \sum_{\tilde{n}=0}^\infty \frac{g^{\tilde{n}}}{\sqrt{\tilde{n}!}}\ket{\tilde{n}}_{n,\text{ph}} \otimes_{m \neq n} \ket{\emptyset}_{m,\text{ph}},
\end{align}
where $\ket{g}$ is a coherent state, and the tilde emphasizes the bosonic bath's number basis instead of the site occupation basis.
Given that
\begin{align}
    UHU^\dagger = \sum_n -g^2\omega_0 \ketbra{n}{n} + \omega_0 b^\dagger_n b_n,
\end{align}
we have
\begin{align}
    C_k(t) &=\frac{1}{Z_B} \sum_{nn'} \frac{e^{-ik(n-n')}}{N} \bra{\emptyset}_{m \neq n,\text{ph}}  \bra{g}_{n,\text{ph}}  e^{-it\sum_n \omega_0 b^\dagger_n b_n} \ket{g}_{n',\text{ph}} \ket{\emptyset}_{m' \neq n',\text{ph}} \bra{n} e^{-it\sum_n -g^2\omega_0 \ket{n}\bra{n}} \ket{n'}\\
&=e^{-g^2} \sum_{n=0}^\infty \frac{g^{2n}}{n!}e^{-it\omega_0 n} e^{itg^2\omega_0},
\end{align}
and therefore,
\begin{align}
    A(\omega)= e^{-g^2} \sum_{n=0}^\infty \frac{g^{2n}}{n!} \delta \left[\omega+g^2\omega_0-\omega_0 n \right].
\end{align}
\subsection{Mean Field Ehrenfest (MFE)}

Given that,
\begin{align}
    \frac{1}{\sqrt{2\omega_0}}(b^\dagger_n+b_n) &\rightarrow Q_n,\\
    i \sqrt{\frac{\omega_0}{2}} (b^\dagger_n-b_n)&\rightarrow P_n,
\end{align}
the mixed quantum-classical Hamiltonian at $J=0$ is given by
\begin{alignat}{2}
    H &= \sum_n  g \sqrt{2\omega_0^3} Q_n \ketbra{n}{n} + &&\frac{1}{2}\omega_0^2 Q_n^2 + \frac{1}{2} P_n^2. \label{eqn:mqchamiltonian}
\end{alignat}
In MFE, to compute $C_k(t)$, we start from the initial state $\ket{\psi(0)}=\ket{k}$, propagate this wavefunction to time $t$ by $H_S+H_{SB}$, and take the overlap $\inner{k}{\psi(t)}$. Since the Hamiltonian is diagonal in the site occupation basis, it is convenient to rewrite the wavefunction in the site basis as
\begin{align}
    \ket{\psi(t)}&=\sum_n \frac{e^{ikn}}{\sqrt{N}} e^{-ig\sqrt{2\omega_0^3} \int_0^t Q_n(t')dt' } \ket{n}.
\end{align}
As a consequence, the equations of motion for the bath variables are
\begin{align}
    \dot{P}_n&=-\bra{\psi(t)}\frac{\partial V}{\partial Q_n}\ket{\psi(t)}-\frac{\partial H_\text{ph}}{\partial Q_n}\\
    &=-\frac{g\sqrt{2\omega^3_0}}{N}-\omega^2_0 Q_n, \label{eqn:p_evol} \\
    \dot{Q}_n&=P_n. \label{eqn:q_evol}
\end{align}
Solving Eq.~\eqref{eqn:p_evol} and Eq.~\eqref{eqn:q_evol} for given initial position and momentum $Q_0$ and $P_0$, we obtain
\begin{align}
    Q_n(t)=\left(Q_0+\sqrt{\frac{2}{\omega_0}} \frac{g}{N}\right) \cos \omega_0 t+\frac{P_0}{\omega_0} \sin \omega_0 t-\sqrt{\frac{2}{\omega_0}} \frac{g}{N}.
\end{align}
Since $Q_n(t)$ does not depend on $n$, we drop the $n$ index. Integrating $Q$ over $dt$ gives
\begin{align}
    \int_0^t Q(t) d t
& =\frac{1}{\omega_0}\left(Q_0+\sqrt{\frac{2}{\omega_0}} \frac{g}{N}\right) \sin \omega_0 t-\frac{P_0}{\omega_0^2} \cos \omega_0 t+\frac{P_0}{\omega_0^2}-\sqrt{\frac{2}{\omega_0}} \frac{g}{N} t.
\end{align}
At zero temperature, $P_0$ and $Q_0$ are sampled from the Wigner function $\rho_W(P_0,Q_0)=\frac{1}{\pi} \exp \left[-P_0^2 / \omega_0-Q_0^2 \omega_0\right]$. Therefore, the correlation function has the form of
\begin{align}
    C_k(t)&=\braket{k}{\psi(t)}\\
    &=\sum_n \frac{1}{N\pi}\int dP_0 dQ_0e^{ -P_0^2 / \omega_0-Q_0^2 \omega_0}e^{-ig\sqrt{2\omega_0^3} \int_0^t Q(t')dt' } \nonumber \\
    &=e^{-g^2} \exp \left(g^2 \cos \omega_0 t\right) \exp \left(-\frac{2 i g^2}{N} \sin \omega_0 t\right) \exp \left(\frac{2 i g^2}{N} \omega_0 t\right). \label{eqn:mfecorr}
\end{align}
Although the correlation function in the $J=0$ Holstein model is expected to be independent of $N$, MFE produces an $N$-dependent correlation function.
This arises because the initial state $\ket{k}$ is formed from a superposition of populations on all $N$ sites.
For the special case of $N=2$ , Eq.~\eqref{eqn:mfecorr} reduces to
\begin{align}
C_k(t)
& =\exp \left(-g^2\right) \exp \left(g^2 e^{-i \omega_0 t}\right) \exp \left(i g^2 \omega_0 t\right),
\end{align}
and the spectral function is
\begin{align}
    A(\omega)= e^{-g^2} \sum_{n=0}^{\infty} \frac{g^{2 n}}{n!} \delta\left[\omega+g^2 \omega_0-\omega_0 n\right].
\end{align}
Thus, when $N=2$, MFE gives the exact results for the Holstein model in the limit of $J=0$.

For general $N$, by the Jacobi-Anger expansion, 
\begin{align}
    e^{z \cos \theta}&=\sum_{n=-\infty}^{\infty} I_n(z) e^{i n \theta},\\
    e^{i z \sin \theta}&=\sum_{n=-\infty}^{\infty} J_n(z) e^{i n \theta},
\end{align}
where $J_n$ and $I_n$ are $n$-th Bessel function and modified Bessel function of the first kind, respectively. Eq.~\eqref{eqn:mfecorr} becomes
\begin{align}
    C_k(t)
    & =\exp \left[-g^2+2 i \frac{g^2}{N} \omega_0 t\right] \sum_{n=-\infty}^{\infty} I_n\left(g^2\right) e^{i n \omega_0 t} 
\sum_{n'=-\infty}^{\infty} J_{n^{\prime}}\left(-\frac{2}{N} g^2\right) e^{i n^{\prime} \omega_0 t},
\end{align}
and the spectral function can be expressed as
\begin{align}
    A(\omega)= e^{-g^2} \sum_{n=-\infty}^{\infty} \sum_{m=-\infty}^{\infty} I_m\left(g^2\right) J_{n-m}\left(-\frac{2}{N} g^2\right) \delta\left[\omega+2 \frac{g^2}{N} \omega_0+n \omega_0\right]. \label{eqn_supp:spec_MFE_generalN}
\end{align}
At $N=2$, Eq.~\eqref{eqn_supp:spec_MFE_generalN} reduces to the exact spectral function:
\begin{align}
    A(\omega)&= e^{-g^2} \sum_{n=-\infty}^{\infty} \sum_{m=-\infty}^{\infty} I_m\left(g^2\right) J_{n-m}\left(- g^2\right) \delta\left[\omega+g^2 \omega_0+n \omega_0\right] \nonumber \\
    &=e^{-g^2}\int_{-\infty}^\infty \sum_{n=-\infty}^{\infty} \sum_{m=-\infty}^{\infty} I_m\left(g^2\right) J_{n-m}\left(- g^2\right)  e^{i\omega t}e^{ig^2 \omega_0 t}e^{in \omega_0 t}dt\\
    &=e^{-g^2}\int_{-\infty}^\infty \sum_{n=-\infty}^{\infty} \sum_{m=-\infty}^{\infty} I_m\left(g^2\right) e^{im \omega_0 t} J_{n-m}\left(- g^2\right) e^{i(n-m) \omega_0 t}  e^{i\omega t}e^{ig^2 \omega_0 t}dt\\
    &=e^{-g^2}\int_{-\infty}^\infty  e^{g^2\cos(\omega_0 t)} e^{-ig^2\sin(\omega_0 t)}   e^{i\omega t}e^{ig^2 \omega_0 t}dt\\
    &=e^{-g^2} \sum_{n=0}^\infty \frac{g^{2n}}{n!} \int_{-\infty}^\infty  e^{-in\omega_0 t}   e^{i\omega t}e^{ig^2 \omega_0 t}dt\\
    &=e^{-g^2} \sum_{n=0}^\infty \frac{g^{2n}}{n!} \delta \left[ \omega+g^2\omega_0-n\omega_0 \right].
\end{align}

\begin{figure}
    \centering
    \includegraphics[width=1.0\linewidth]{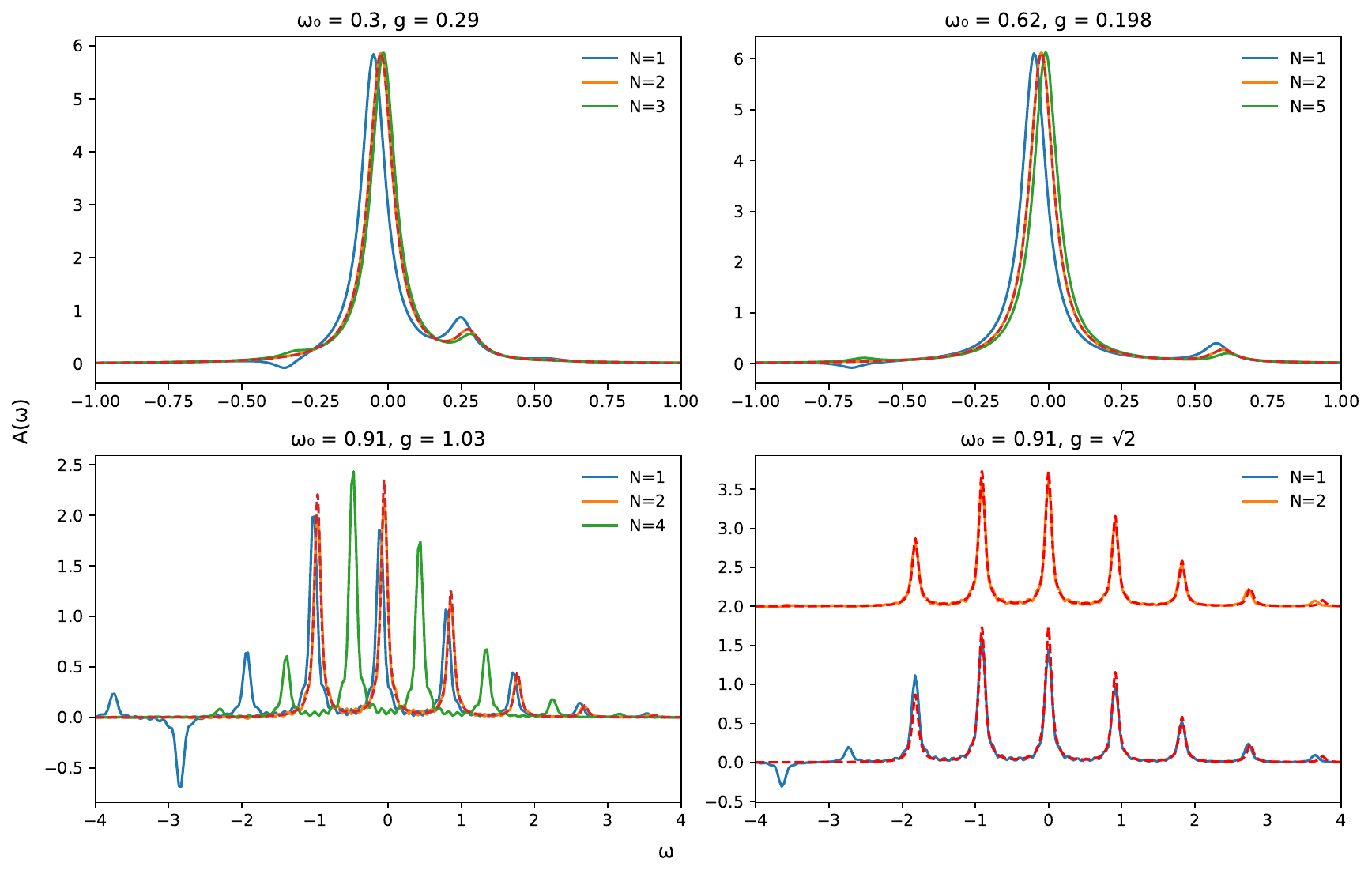}
    \caption{Spectral functions computed by MFE for $J=0$ at zero temperature. Each panel represents a unique set of coupling strength $g$ and bath frequency $\omega_0$ across different lattice sizes. The red dashed lines represent the exact solution.}
    \label{fig:mfeJ=0}
\end{figure}

As shown in Fig.~\ref{fig:mfeJ=0}, the spectral function computed by MFE is dependent on the number of sites $N$, with $N=2$ always producing exact results. When value of $g^2$ is such that $g^2(\frac{2}{N}-1)$ is an integer (as seen in the lower right panel of Fig.~\ref{fig:mfeJ=0}), MFE for any $N$ will generate peaks that align with the exact solution. However, for $N \neq 2$, MFE introduces additional peaks that are not present in the exact results.

\subsection{Multi-State Mapping Approach to Surface Hopping (MASH)}
Given the mixed quantum-classical Hamiltonian at $J=0$ in Eq.~\eqref{eqn:mqchamiltonian}, it is evident that the Hamiltonian is diagonal in the site occupation basis. As a result, the adiabatic surfaces correspond directly to the site occupation states. The active surface, $\ket{\alpha}$, represents one of them and does not change during the time evolution. We can derive the equations of motion for the bath variables associated with this active site as
\begin{align}
\dot{P}_\alpha & =-\left\langle\alpha\left|\frac{\partial V}{\partial Q_\alpha}\right| \alpha\right\rangle-\frac{\partial H_{\text{ph}}}{\partial Q_\alpha} \nonumber \\
& =-g \sqrt{2 \omega_0^3}-\omega_0^2 Q_\alpha, \label{Pa_evol} \\
\dot{Q}_\alpha & =P_\alpha.\label{Qa_evol}
\end{align}
For other sites not the active surface, $n \neq \alpha$,
\begin{align}
\dot{P}_n&=-\left\langle\alpha\left|\frac{\partial V}{\partial Q_n}\right| \alpha\right\rangle-\frac{\partial H_{\text{ph}}}{\partial Q_n} \nonumber \\
&=-\omega_0^2 Q_n , \label{Pn_evol} \\
\dot{Q}_n&=P_n. \label{Qn_evol}
\end{align}
In MASH, the correlation function is expressed as
\begin{align}
    C_k(t)=\int d\vec{P}_0 d\vec{Q}_0 \rho_W(\vec{P}_0,\vec{Q}_0, \beta) \int d\vec{c}  \rho_k(\vec{c}) W_{kk}(t).
\end{align}
To compute the correlation function $C_k(t)$, we use focused sampling to sample the initial wavefunction as follows,
\begin{align}
    \ket{\psi(0)}=\sqrt{\frac{1+\beta_{N+1}}{\alpha_{N+1}}}e^{i\theta_{\tilde{k}}} \ket{\tilde{k}}+\sqrt{\frac{\beta_{N+1}}{\alpha_{N+1}}} e^{i\theta_{\tilde{\phi}}} \ket{\tilde{\phi}}+\sum_{\substack{q\neq \tilde{k} \\ q\neq \tilde{\phi}}} \sqrt{\frac{\beta_{N+1}}{\alpha_{N+1}}} e^{i\theta_{q}} \ket{q},
\end{align}
where
\begin{align}
    \ket{\tilde{k}}&=m_k \ket{k} + m_\phi \ket{\phi},\\
    \ket{\tilde{\phi}}&=m_\phi \ket{k} - m_k \ket{\phi},
\end{align}
and $m_k,~m_\phi$ are real numbers. Note the use of $N+1$ in the $\alpha$ and $\beta$ coefficients to emphasize that the auxiliary state must be included here, and $N$ is the size of the original system \textit{without} the auxiliary state.
The wavefunction can be rewritten in the momentum basis as
\begin{align}
|\psi(0)\rangle
&=\left(m_\phi\sqrt{\frac{1+\beta_{N+1}}{\alpha_{N+1}}} e^{i \theta_{\tilde{k}}}-m_k\sqrt{\frac{\beta_{N+1}}{\alpha_{N+1}}} e^{i \theta_{\tilde{\phi}}}\right)|\phi\rangle +\left(m_k\sqrt{\frac{1+\beta_{N+1}}{\alpha_{N+1}}} e^{i \theta_{\tilde{k}}}+m_\phi\sqrt{\frac{\beta_{N+1}}{ \alpha_{N+1}}} e^{i \theta_{\tilde{\phi}}}\right)\ket{k} +\sum_{q} \sqrt{\frac{\beta_{N+1}}{\alpha_{N+1}}} e^{i \theta_{q}}|q\rangle. \label{psi0}
\end{align}
Since the Hamiltonian is diagonal in real space, we rewrite the wavefunction in the site occupation basis and propagate it in time to find
\begin{align}
|\Psi(t)\rangle 
&=\left(m_\phi\sqrt{\frac{1+\beta_{N+1}}{ \alpha_{N+1}}} e^{i \theta_{\tilde{k}}}-m_k\sqrt{\frac{\beta_{N+1}}{ \alpha_{N+1}}} e^{i \theta_{\tilde{\phi}}}\right)|\phi\rangle+\sum_n e^{-i g \sqrt{2 \omega_0^3} \int_0^t Q_n(t') d t'} \nonumber \\
& \left(m_k\sqrt{\frac{1+\beta_{N+1}}{ \alpha_{N+1} N}} e^{i \theta_{\tilde{k}}} e^{-i k n}+m_\phi\sqrt{\frac{\beta_{N+1}}{ \alpha_{N+1} N}} e^{i \theta_{\tilde{\phi}}} e^{-i k n}+\sum_{q} \sqrt{\frac{\beta_{N+1}}{\alpha_{N+1} N}} e^{i \theta_{q} }e^{-i q n}\right)|n\rangle. \label{eq:MASHwf_inrspace}
\end{align}
In the momentum basis, the wavefunction becomes
\begin{align}
|\psi(t)\rangle &=\left(m_\phi\sqrt{\frac{1+\beta_{N+1}}{ \alpha_{N+1}}} e^{i \theta_{\tilde{k}}}-m_k\sqrt{\frac{\beta_{N+1}}{ \alpha_{N+1}}} e^{i \theta_{\tilde{\phi}}}\right)|\phi\rangle \nonumber \\
&+\sum_{n}\frac{1}{N}e^{-i g \sqrt{2 \omega_0^3} \int_0^t Q_n(t') d t'}\sum_k \left(m_k\sqrt{\frac{1+\beta_{N+1}}{ \alpha_{N+1} }} e^{i \theta_{\tilde{k}}} +m_\phi\sqrt{\frac{\beta_{N+1}}{ \alpha_{N+1} }} e^{i \theta_{\tilde{\phi}}} +\sum_{q } \sqrt{\frac{\beta_{N+1}}{\alpha_{N+1} }} e^{i \theta_{q} } e^{i(k-q)n} \right)|k\rangle. \label{psi_t}
\end{align}
Plugging in the coefficients $c_k(0)$ from Eq.~\eqref{psi0} and $c_k(t)$ from Eq.~\eqref{psi_t}, and integrating over the phases, we find
\begin{align}
    C_k(t)&=\frac{1}{\pi}\int dP_0 dQ_0 e^{-P^2_0/\omega_0} e^{-Q^2_0\omega_0} \int_0^{2\pi} d\vec{\theta} \frac{c_k(\vec{\theta},P_0,Q_0,t)}{c_k(\vec{\theta},t=0)} \\
    &=\sum_n \int d P_0 d Q_0 e^{-P_0^2 / \omega_0} e^{-Q_0^2 \omega_0} \frac{e^{-i g \sqrt{2 \omega_0^3} \int_0^t Q_n(t') d t'}}{\pi N}
\end{align}
It is evident that $C_k(t)$ and the corresponding spectral function are expressed as a sum of two components: one dependent on $Q_\alpha(t)$, and the other on $Q_n(t)$, where $n$ refers to a site that is not the active surface which remains unchanged throughout the propagation.

\subsubsection{Spectral function for state $\ket{\alpha}$}
Solving Eq.~\eqref{Pa_evol} and Eq.~\eqref{Qa_evol} gives
\begin{align}
    Q_\alpha(t)& =\left(Q_0+\sqrt{\frac{2}{\omega_0}} g\right) \cos \omega_0 t+\frac{P_0}{\omega_0} \sin \omega_0 t-\sqrt{\frac{2}{\omega_0}} g 
\end{align}
Therefore,
\begin{align}
C_{\alpha, k}(t) & =\frac{1}{N} \exp \left[-g^2 \right] \exp \left[g^2 \cos \omega_0 t\right]  \exp \left[-i 2 g^2 \sin \omega_0 t\right] \exp \left[i 2 g^2 \omega_0 t\right],
\end{align}
and
\begin{align}
    A_{\alpha, k}(\omega) =\frac{1}{N} e^{-g^2} \sum_{n=-\infty}^{\infty} \sum_{m=-\infty}^{\infty} I_m\left(g^2\right) J_{n-m}\left(-2 g^2\right)\delta\left[\omega+2 g^2 \omega_0+\omega_0 n\right], \label{spectral_alpha}
\end{align}
where $J_n$ and $I_n$ are the $n^{th}$ Bessel function and modified Bessel function of the first kind, respectively.

\subsubsection{Spectral function for state $|n \neq \alpha \rangle$}
Similarly, solving Eq.~\eqref{Pn_evol} and Eq.~\eqref{Qn_evol} gives
\begin{align}
    Q_n(t)=Q_0 \cos \omega_0 t+\frac{P_0}{\omega_0} \sin \omega_0 t,
\end{align}
which leads to
\begin{align}
C_{n, k}(t)=\frac{N-1}{N} e^{-g^2} \exp \left[g^2 \cos \omega t\right]
\end{align}
and
\begin{align}
    A_{n, k}(\omega)= \frac{N-1}{N} e^{-g^2} \sum_{n=-\infty}^{\infty} I_n\left(g^2\right) \delta[\omega+\omega_0n]. \label{spectral_n}
\end{align}

\begin{figure}
    \centering
    \includegraphics[width=1.0\linewidth]{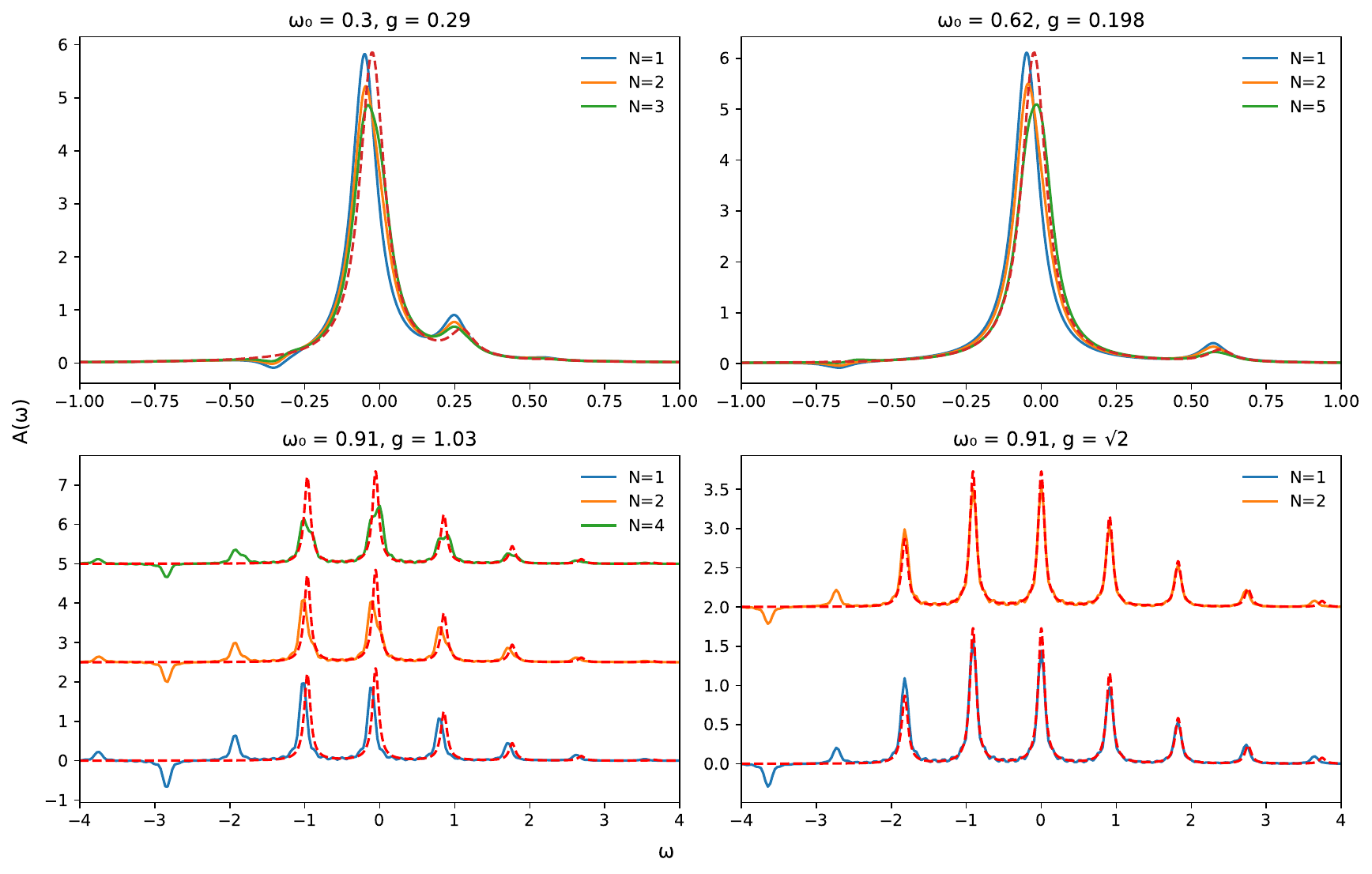}
    \caption{Spectral functions computed by MASH for $J=0$ at zero temperature. Each panel represents a unique set of coupling strength $g$ and bath frequency $\omega_0$ across different lattice sizes. The red dashed lines represent the exact solution.}
    \label{fig:mashJ=0}
\end{figure}

Fig.~\ref{fig:mashJ=0} illustrates that MASH fails to produce accurate results across all parameter regimes. Although it generates peaks that coincide with the exact solution when $g^2$ is an integer (as shown in the lower right panel of Fig.~\ref{fig:mashJ=0}), it also introduces additional peaks that are absent in the exact results. When $g^2$ is not an integer and $N>1$, two distinct series of peaks emerge: one originating from the active site, as described in Eq.~\eqref{spectral_alpha}, and the other from the remaining sites, as described in Eq.~\eqref{spectral_n}. As $N$ increases, the peaks from the remaining sites grow in weight, leading to a broadening effect evident in the upper panels and lower left panel of Fig.~\ref{fig:mashJ=0}.

\section{Variational Monte Carlo with neural quantum states}
Here we provide a brief description of the VMC method with a NN ansatz. We refer the reader to Ref. \cite{mahajan2024structure} for more details. The NN wave function is given by
\begin{equation}
   \ket{\psi} = \sum_{\textbf{n}}\frac{\exp\left[f(\textbf{n})\right]}{\sqrt{\prod_i \nu_i!}}\ket{\mathbf{n}},\label{eq:nn_wave_function}.
\end{equation}
Here \(\textbf{n} = (\left\{e_i\right\}, \left\{\nu_i\right\})\) is a configuration, where \(\left\{e_i\right\}\) and \(\left\{\nu_i\right\}\) are electron and phonon occupation numbers, respectively. The function \(f\) is a sum of two neural nets:
\begin{equation}
   f(\textbf{n}) = r(\textbf{n}) + i\phi(\textbf{n}).
\end{equation}
We use multilayer perceptron NNs with real-valued weights and biases in this work. Properties of this wave function, such as the energy, are evaluated using Monte Carlo sampling. The wave function parameters are optimized using a generalized gradient descent method to minimize the energy.

We calculate spectral functions by working in the tangent space of the optimized ground state wave function (\(\ket{\psi_0}\)) with basis functions given by
	\begin{equation}
		\ket{\psi_{\mu}} = \frac{\partial \ket{\psi_0}}{\partial p_{\nu}}.
	\end{equation}
where \(p_{\mu}\) are the optimized variational parameters. The resulting basis is nonorthogonal and has linear dependencies. Thus to obtain the excited states and spectral functions, we solve the generalized eigenvalue problem
\begin{equation}
   \mathbf{H}\mathbf{C} = E\mathbf{S}\mathbf{C},
\end{equation} 
where the Hamiltonian and overlap matrices are given by
\begin{equation}
   \begin{split}
      \mathbf{H}_{\mu\nu} &= \expecth{\psi_{\mu}}{H}{\psi_{\nu}},\\
      \mathbf{S}_{\mu\nu} &= \inner{\psi_{\mu}}{\psi_{\nu}}.
   \end{split}\label{eq:lr_vmc}
\end{equation}
We use the eigenspectrum of this effective Hamiltonian \(\left\{E_{i}, \ket{i}\right\}\) to calculate the one-particle spectral function at zero temperature as
 \begin{equation}
   \begin{split}
       A(k, \omega) &= -\frac{1}{\pi}\text{Im}\langle 0| c_k \frac{1}{\omega - H + i\eta} c^{\dagger}_k|0\rangle\\
       &= -\frac{1}{\pi}\sum_{i}|\expecth{0}{c_k}{i}|^2 \frac{\eta}{(\omega-E_i)^2+\eta^2},\\
   \end{split}
\end{equation}
where \(\eta\) is a Lorentzian broadening parameter.

%